\documentclass[prd]{revtex4}
\usepackage{epsfig}
\usepackage{mathrsfs}
\usepackage{amsmath}
\usepackage{amssymb}
\usepackage{color}
\usepackage{graphicx}
\newcommand{\rig}{\mathcal{R}}
\newcommand{\flur}{J}

\begin{document}

\title{On the ``Hysteresis Effects'' observed by AMS02 in Cosmic Ray Solar Modulations}

\author{Paolo Lipari}
\email{paolo.lipari@roma1.infn.it}
\affiliation{INFN, Sezione Roma ``Sapienza'',
 piazzale A.Moro 2, 00185 Roma, Italy}

\author{Silvia Vernetto}
\email{vernetto@to.infn.it}
\affiliation{
INAF, Osservatorio Astrofisico di Torino,
 via P.Giuria 1, 10125 Torino, Italy}

\affiliation{
 INFN, Sezione Torino,
 via P.Giuria 1, 10125 Torino, Italy}

\begin{abstract}
The AMS02 collaboration has recently published high precision daily measurements of the spectra of cosmic ray protons, helium nuclei and electrons taken during a time interval of approximately 10 years from 2011 to 2020. Positron spectra averaged over distinct 27 days intervals have also been made public. The AMS02 collaboration has shown some intriguing "hysteresis" effects observed comparing the fluxes of protons and helium nuclei or protons and electrons. In this work we address the question of the origin of these effects. We find that the spectral distortions generated by propagation in the heliosphere are significantly different for particles with electric charge of opposite sign (an effect already well established), with different behaviour before and after the solar magnetic field polarity reversal at solar maximum. This results in hysteresis effects for the $p/e^-$ comparison that follow the 22--year solar cycle. On the other hand particles with electric charge of the same sign suffer modulations that are approximately equal. The hysteresis effects observed for a helium/proton comparison can then be understood as the consequence of the fact that the two particles have interstellar spectra of different shape, and the approximately equal spectral distortions generated by propagation in the heliosphere have a rigidity dependence that is a function of time. These hysteresis effects can in fact be observed studying the time dependence of the shape of the spectra of a single particle type, and also generate short time loop--like structures in the hysteresis curves correlated with large solar activity events such as coronal mass ejections (CME's). A description of solar modulations that includes these effects must go beyond the simple Force Field Approximation (FFA) model. A minimal, two--parameter generalization of the FFA model that gives a good description of the observations is presented.
\end{abstract}

\maketitle

\section{Introduction}
\label{sec:introduction} 
The time dependence of the fluxes of Galactic cosmic rays generated by 
solar modulations \cite{Potgieter:2013pdj} has been studied for several decades.
During most of this time, the fundamental instrument to study these effects 
has been the neutron monitor \cite{simpson}, but in recent years 
the
PAMELA \cite{Adriani:2013as,Martucci:2018pau,Adriani:2015kxa,Adriani:2016uhu}
and AMS02 \cite{ams_bartels_protons,ams_bartels_electrons,ams_daily_protons,ams_daily_helium,ams_daily_electrons,ams_helium_isotopes}
detectors, located on satellites, 
have obtained precise direct measurements of the CR spectra
that allow much more detailed analysis.

The AMS02 collaboration has published
measurements of the spectra for four different particle types
(protons, helium nuclei, electrons and positrons)
averaged in time during 79 Bartels rotation of the Sun
(each lasting 27 days) \cite{ams_bartels_protons,ams_bartels_electrons},
and more recently daily spectra
for protons
\cite{ams_daily_protons}, 
helium nuclei
\cite{ams_daily_helium}
and electrons
\cite{ams_daily_electrons}
that extend
for several years: 2824 spectra taken during a time period
of 8.44~yr for $p$ and He, and 3193 spectra taken
during a period of 10.45~yr for $e^-$,
with both data sets starting on 2011-05--20.
These data contain an enormous amount of information about the
dynamics of the heliosphere and the properties of propagation
of relativistic charged
particles in it, and are the object of multiple studies.

In their most recent papers the AMS02 collaboration has discussed
some intriguing ``hysteresis effects'' observed comparing the time dependence
of the fluxes for different particle types. 
In \cite{ams_daily_helium} the ratio of helium and proton
fluxes in some fixed rigidity ranges
is studied as a function of the helium flux.
Comparing moving averages of the two quantities
with an integration time interval of 378 days (14 Bartels rotations)
and one day step, the authors find that one
value of the helium flux does not correspond to a unique value of the He/$p$ ratio
(and therefore to a unique value of the proton flux).
The time averaged He/$p$ ratio is found to be higher after solar maximum,
and the authors conclude that at low rigidity the modulation
of the helium to proton flux ratio
is different before and after the solar maximum of 2014.

In \cite{ams_daily_electrons} a similar study is performed
for electron and proton spectra, comparing the time
dependence of the two fluxes in the same rigidity intervals. 
Also in this case it is observed that one value 
of the $p$ flux does not correspond to a unique
value of the $e^-$ flux.
For long averaging time intervals
(such as $T = 378$~days or 14 Bartels rotations) one observes that, 
for the same proton flux, the electron flux is significantly
smaller after solar maximum. The effect is similar
to the one observed comparing the helium and proton spectra,
but it is one order of magnitude larger.
The study of moving averages of the fluxes with shorter integration times
reveals additional structures in the time dependence
for the $e^-/p$ ratio that
appears to be associated to the presence of
transients of solar activity, that are also the cause of rapid
time variations of the fluxes of both particles.
Studies of moving averages of the He/$p$ ratio with shorter time intervals 
have not been discussed in the AMS02 publications, but also this 
ratio exhibits time structures similar to those observed for
the $e^-/p$ case. 

In the following we want to address the problem of the origin
of the ``hysteresis'' effects observed by AMS02. We will
show that two essentially different mechanisms are operating.
One mechanism is relevant for the long time scale dependence of the
$e^-/p$ ratio, and has its origin in the well established fact
that CR particles with electric charge of opposite sign
travel along different trajectories that are confined in different
regions of the heliosphere, and this results in different modulations effects.
The heliospheric trajectories depend on the polarity of the solar
magnetic field, and the reversal of the polarity at solar maximum 
is the origin of the large differences in the $e^-/p$ ratio before and after
the solar maximum of 2014.

A second more subtle physical mechanism is at the origin of the hysteresis effects
observed for the He/$p$ ratio,
In this case the solar modulations for the two particle types
are (in a sense that will be made more precisely below) in good approximation equal,
and the hysteresis effects are the result of the fact that
the spectral distortions generated by the modulations
can have different rigidity dependences at different times.

This second mechanism can be observed, and is in fact more easily understood,
studying the time dependence of the fluxes of one single particle type in
two distinct rigidity intervals. The point is that 
observations of the same value of the flux at rigidity $\rig_1$ can correspond
(for different observations times) to different values of the flux at
the rigidity $\rig_2$. Therefore the plot of one flux versus the other 
[$J(\rig_2, t)$ versus $J(\rig_1, t)$] can exhibit non trivial ``hysteresis''
structures.

This mechanism generates similar effects also in the comparison
of the fluxes $J_p(\rig, t)$ and $J_{\rm He} (\rig,t)$ 
of protons and helium at the same value of rigidity,
even if the spectra of the two particles suffer the same modulations.
This is because the effects of modulations must 
be understood not as an energy (or rigidity) dependent absorption effect,
but instead as a distortion that acts on the local interstellar (LIS) spectra,
and depends not only on state of the heliosphere, but also on the shape
of the LIS spectra, that are different for protons and helium nuclei.

A simplified way to understand and model the solar modulations
is to describe them as the effect of an average energy loss $\Delta E$
suffered by particles during propagation in the heliosphere.
The hysteresis effects observed comparing the spectra of protons
and electrons are due
to the fact that the heliospheric energy losses for
$p$ and $e^-$ are different and change in different ways during the solar cycle.
On the contrary, the hysteresis effects observed comparing the spectra
of protons and helium nuclei are due to the fact the their
heliospheric energy losses 
(that have in good approximation the same time
and rigitity dependences being related by the simple equation
$\Delta E_{\rm He} = 2 \, \Delta E_p$)
have a non trivial rigidity dependence that takes different shapes at different times.

This paper is organized as follows.
In the next section we show how ``hysteresis effects'' are present 
in the AMS02 daily spectra measurements of all three particles
(protons, helium nuclei and electrons) and can be observed studying
the fluxes of each single particle type, with no need
to compare different particle types.
In the following section 
we introduce a very simple parametrization for the rigidity (or energy) spectra,
that can describe surprisingly well the data for $p$, He, and $e^\mp$.
This parametrization has two time independent parameters:
a normalization and a spectral index that together define 
a simple power law in rigidity,
and two time dependent parameters that determine a
rigidity dependent potential that controls the modulation effects.
Section~\ref{sec:potentials} presents the time dependence of the potentials
for the different particles during the extended time interval
of the PAMELA and AMS02 observations.
Section~\ref{sec:lis} discuss the physical meaning of the modulation
potential we have introduced,  and the shape of the local interstellar (LIS)
spectra of the CR particles.
Section~\ref{sec:hysteresis} discusses the  ``loops'' of  different periods
that  emerge from  different hysteresis studies.
The  final section summarizes the results.

\section{Flux correlations for a single particle type}
\label{sec:flux_corr}
The effect we want to investigate here is the {\em shape} of the
distortions generated by solar modulations on the
rigidity (or energy) spectrum of one particle type at different times.
It is well known that at low rigidity the CR fluxes are time
dependent, and for example the proton flux at
$\rig \simeq 1$~GV changes being highest
(lowest) at the minimum (maximum) of solar activity.
The question we want to address is
if the value of the $p$ flux at 1~GV
determines the entire spectrum at all rigidities or not.
This is in fact the case in models 
that describe solar modulations
in terms of only one time dependent parameter, such as the commonly used
Force Field Approximation (FFA) \cite{Gleeson:1968zza},
where a measurement of the flux at one rigidity 
(if it is in the range where the effects of the modulations are not negligible)
is sufficient to determine the entire spectrum.

\begin{figure}
\begin{center}
\includegraphics[width=14cm]{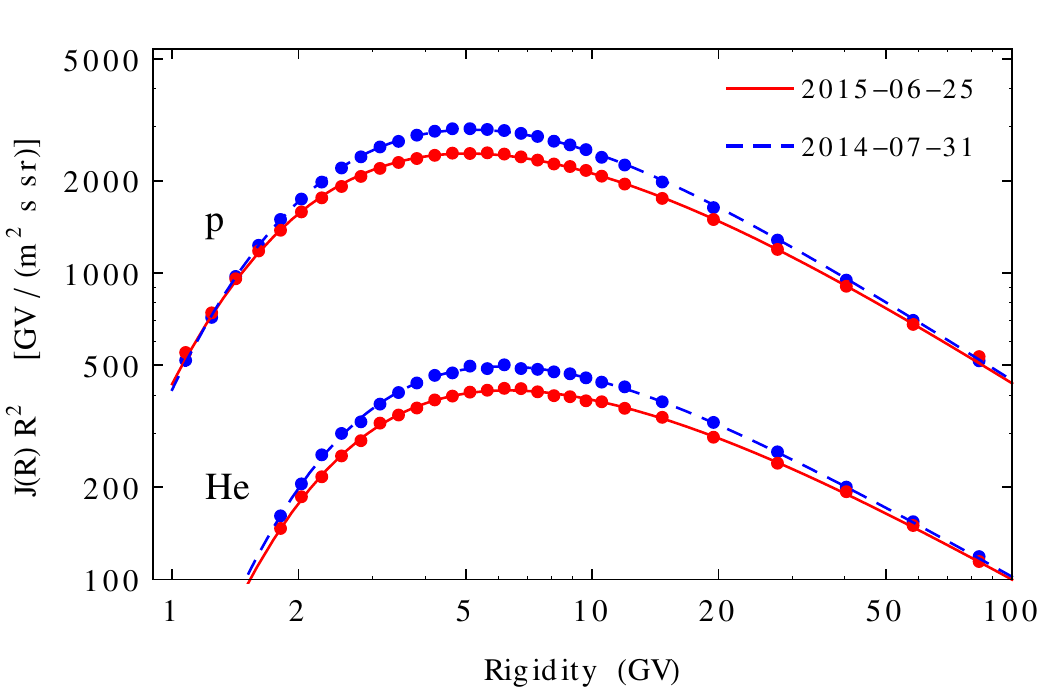}
\end{center}
\caption {\footnotesize Rigidity spectra for protons and
 helium nuclei observed during two different days
 (2014--07-31 and 2015--06--25) by AMS02 \cite{ams_daily_protons,ams_daily_helium}.
 The lines are fits to the spectra discussed in the main text.
 \label{fig:flux_example}}
\end{figure}

The AMS02 data however show that the
assumption that solar modulation can be described
by a single time dependent parameter is not correct.
This conclusion emerges directly from the data, without any analysis.
An illustration of this is presented in Fig.~\ref{fig:flux_example}
that shows the proton spectra
measured by AMS02 \cite{ams_daily_protons} during two different
days (2014--07--31 and 2016-06--25).
The average fluxes measured during these two days
are approximately equal for a rigidity of order 1~GeV,
but differ by ($20\pm 1)$~\% in the rigidity bin 
[4.88--5.37]~GV.
Fig.~\ref{fig:flux_example} also shows the spectra of
helium nuclei measured by AMS02 \cite{ams_daily_helium}
during the same two days.
One can note that the
effects of solar modulations for protons and helium nuclei
have the same qualitative features, as 
also the helium spectra are approximately equal ar $\rig \simeq 1$~GV,
and differ by $\sim 20$\% at $\rig \simeq 5$~GV.
A more quantitative study, presented below,
will show that the distortions to the proton and helium spectra
(and in fact also to the positron spectrum) 
generated by solar modulations are in fact in very good approximation equal.

The observation that the  flux at one rigidity $\rig_1$ can correspond,
at different times, to different fluxes at a second rigidity $\rig_2$,
suggests to explore the possibility to observe ``hysteresis'' effects
such as those discussed by AMS02 (for the fluxes of two different particles
measured in the same rigidity interval) \cite{ams_daily_helium,ams_daily_electrons},
also for the fluxes of one single particle type in two distinct rigidity intervals.

Some results of this type of study are illustrated in
Fig.~\ref{fig:corr_particles}. The three panels in the top row of
the figure show the time dependence of the flux of
protons \cite{ams_daily_protons},
helium nuclei \cite{ams_daily_helium}
and electrons \cite{ams_daily_electrons}
measured during different days in one fixed interval of rigidity
($\rig =[1,1.16]$~GV for $p$, 
$[1.71,1.92]$~GV for He and 
$[1,1.71]$~GV for $e^-$).
The measurements for protons and helium are taken
for 2717 different days \footnote{The AMS02 data has released
 2824 proton and helium spectra, but in 107 of them the
 measurements are given only for $\rig > 2.97$~GV.}
from 2011-05-20 to 2019--10--29, 
while the measurements for electrons are taken for 3193 days
from the same initial day and extending to 2021--11-02.

The time interval of the daily spectra
measurements covers a large part of
the 24th solar cycle that extends from the minimum in December 2008 to
the next minimum in December 2019 passing through a maximum
around April 2014 (the $e^-$ measurements cover also the beginning
of cycle 25th). 

The fluxes for all three particle
exhibit significant time variations on a variety of time scales,
with the most prominent effect associated to
the 11~year solar cycle.
An important point is to note that, superimposed on
the general trend of decreasing fluxes
before solar maximum and increasing fluxes after maximum,
other significant time variation structures associated to
phases of enhanced or suppressed solar activity are present.
For example, during the first part of the cycle
(increasing solar activity and decreasing CR fluxes) one can
identify three main local maxima of the flux, and during the
second part of the cycle
(decreasing solar activity and increasing CR fluxes)
a prominent minimum of the fluxes is present in September 2017.
In the three plots the colors and the vertical lines identify some
time intervals associated with prominent time structures.
These time intervals are labeled with a letter,
with intervals (a), (b) and (c) roughly centered on
the three local maxima in the first part of the cycle;
intervals (d) and (e) covering the solar maximum part of the cycle;
while the time interval around the prominent local minimum of
September 2017 is labeled (h). The same colors are used in subsequent plots
to identify the same time intervals.

It is interesting to note that while the time dependences
of the flux for the three particles ($p$, He and $e^-$) are
qualitatively similar, the similarity is remarkably accurate
for protons and helium nuclei, while there is an evident
large difference between electrons and the two positively charged
particles. To illustrate this point, the time dependences
for helium nuclei and electrons are shown in the form $\phi(t)/\langle \phi \rangle -1$
(where the average is taken during the same time interval for
all particles), and compared with the time evolution for protons.

The nine panels in the lower part of Fig.~\ref{fig:corr_particles}
show the time evolution of the fluxes 
of protons, helium nuclei and electrons measured
simultaneously in two distinct rigidity intervals.
This is achieved studying the ``trajectory'' in time
of the pair $\{\flur_1(t), \flur_2 (t)\}$, where
$\flur_1(t)$ and $\flur_2 (t)$ are the fluxes measured at time $t$
in the two different rigidity intervals
[$\rig_1, \rig_2]$ and 
[$\rig_1^\prime, \rig_2^\prime ]$.

The three columns in Fig.~\ref{fig:corr_particles}
shows the trajectories of pairs of fluxes for
protons (left), helium nuclei (middle) and electrons (right).
In all three cases the lower rigidity interval is the same
used to show the time evolution of the fluxes in the top row
($\rig =[1,1.16]$~GV for $p$,
$[1.71,1.92]$~GV for He and
$[1,1.71]$~GV for $e^-$),
while the second rigidity intervals are:
$\rig =[2.97,3.29]$~GV for $p$,
$[3.64,4.02]$~GV for He and
$[2.97,4.02]$~GV for $e^-$.

In the second row of panels, the time evolution of the
pair of fluxes $\{J_1(t), J_2 (t)\}$
is shown as a broken line that connects the daily measurements.
There is of course a strong correlation between $J_1(t)$ and
$J_2(t)$. This is expected because both fluxes are large (small)
in periods of weak (strong) solar activity, however
the trajectory is not limited to a narrow band, as expected
if one assumes that the value of the flux $J_2(t)$ is determined by the 
value of $J_1(t)$. The spread of values of $J_2(t)$ for a
fixed value of $J_1(t)$ is larger than the
errors on the measurement
(that are of order 1\%, 1.5\% and 2\% for $p$, He and $e^-$ and are not
shown to avoid cluttering), and therefore is physically significant.

The trajectories that describe the daily flux measurements
have a rich and complex structure that 
encodes very valuable information about CR propagation in the heliosphere,
but because of their complexity are also difficult to interpret,
and for this reason it is interesting to perform
moving averages of the measurements, even if this
procedure erases significant information
about the time evolution of the spectra.

The third row of panels in Fig.~\ref{fig:corr_particles} shows
moving averages of the trajectories $\{J_1(t), J_2(t)\}$
for an averaging time interval of 81 days (3 Bartels rotations)
and one day step.
The panels in the bottom row 
show the same moving averages of the flux pairs
but in a slightly different form, replacing
the value of $\flur_2 (t)$ with its deviation from an 
average value (shown as a dashed line in the two panels above).
The resulting trajectories are much simpler, and
reveal interesting structures in the time evolution of the
spectra that are analogous (and in fact encode the same effects)
of what has been observed by the AMS02 collaboration in the study
of the He/$p$ and $e^-/p$ ratios.

The qualitative feature that is most evident in the figure
is the presence of ``hysteresis loops'' in the trajectories
that trace the evolution of the flux pairs $\{\flur_1(t), \flur_2(t)\}$.
Inspecting Fig.~\ref{fig:corr_particles}
one can identify three such loops during the first part
of the solar cycle (when solar activity is going toward maximum)
that corresponds to the time intervals (a), (b) and (c)
(following the notation indicated in the top row of the figure),
and one loop during the second part of the solar cycle
(when solar activity is decreasing after solar maximum),
and corresponds to the time interval (h).

The ``loops'' are related to strong perturbations of the
interplanetary magnetic field, superimposed
to the more gradual 11 year solar cycle.
The loops in the first part of the cycle are formed
when the general decreasing trend of the two fluxes
$\flur_1(t)$ and $\flur_2(t)$ is inverted
and both fluxes increase during a short time interval
before returning to their normal behavior of gradual decrease.
The effect is faster and relatively larger
for the flux in the high rigidity bin,
generating a clockwise loop in the trajectory.

In the second part of the cycle (after solar maximum)
a prominent loop is present around September 2017,
when some large coronal mass ejections (CME)
generate a large suppression of the CR fluxes during a time interval
of several months. Also in this case the response of the flux
in the high rigidity bin is larger and faster, resulting
again in a clockwise loop in the trajectory of the flux pair.

It is straightforward to see how these loop structures
in the time evolution of the CR spectra are also visible 
comparing the fluxes of two different particle types, 
as done in the AMS02 papers \cite{ams_daily_helium,ams_daily_electrons}.

The study of the rigidity dependence of solar modulations has been studied
for decades, in particular in association with
the so called Forbush decreases, sudden drops of the CR spectra
(associated to CME's or high-speed streams from coronal holes) 
first observed in 1937 \cite{forbush-1937}. 
Most of these studies have been performed with ground--based neutron monitor (NM) detectors.
These instruments are located in regions with different geomagnetic cutoffs,
and therefore can observe CR flux variations integrating over different rigidity ranges.
Comparisons of the counting rates of different NM detectors have allowed to observe
already in the 1970's the presence of ``hysteresis loops'' associated to the
rigidity dependent modulations \cite{hyst1,rajan_loops}.

In more recent times spaceborne detectors placed in near Earth orbit
have been able to measure directly the CR spectra of different particles
(protons, helium nuclei and electrons by PAMELA
\cite{Munini:2018cgc,Lagoida:2021udw}
and electrons and positrons by DAMPE \cite{DAMPE:2021qet})
during major Forbush decreases, obtaining evidence that the CR fluxes
recovery times are rigidity dependent and shorter at higher $\rig$.
The AMS02 data, thanks to its large statistics, high precision and a
an extended data taking is of great value to develop a more complete
understanding of the effects of perturbations in the
interplanetary environment on the CR spectra.

\section{A two--parameter phenomenological description of solar modulations}
\label{sec:parametrization}
The discussion in the previous section, as in the 
papers that present the AMS02 measurements,
has been developed studying the time dependence of directly
measured fluxes.
This approach has the merit of avoiding the introduction of model
dependent quantities and concepts, however it has also significant limitations.
This is in part because it is not ``economic'', since there
are infinite ways to choose the rigidity or energy intervals used to
study the evolution of the spectra, moreover, such a discussion cannot completely capture
the properties of the modulation mechanism that generates distortions to 
the {\em shape} of the CR spectra.

In the following we will attempt to develop a simple parametrization of the
CR spectra with the goal of extracting from the data few quantities that
can capture the main effects of solar modulations. 
A convenient starting point is the widely used
and very successful model of the Force Field Approximation (FFA)
introduced by Gleeson and Axford \cite{Gleeson:1968zza}.
The fundamental assumption in the model is that 
CR particles traversing the heliosphere suffer a time dependent energy loss 
$\Delta E = |q|\, V(t)$ proportional to
the absolute value of their electric charge.
In the original version of the FFA, the same potential $V$ is valid for
all particle types, but it is now well established that particles with
electric charge of opposite sign propagate in different regions of the
heliosphere and therefore ``see'' different potentials.
The question if the same potential can describe the modulations of
all particles that have electric charge of the same sign, should of course
be tested experimentally.

If the LIS spectra at the boundary of the heliosphere are, as expected,
isotropic and constant in time, it is then
straightforward to derive an expression
for the energy spectrum observable at the Earth at time $t$:
\begin{equation}
\phi(E, t) = \frac{p^2}{p_0^2} \; \phi_0 [E + |q| \, V(t)]
\label{eq:ffa}
\end{equation}
In this expression $\phi_0 (E)$ is the LIS spectrum,
and $p$ and $p_0$ are the 3--momenta
that correspond to the energies $E$ and $E + |q| \, V(t)$,
that are the energies of a CR particle when detected at the
Earth and entering the heliosphere.

In the FFA model the solar modulations are calculated
in terms of the LIS spectrum $\phi_0 (E)$, but the validity of the model
can be tested without any knowledge of this spectrum,
simply comparing spectra that are directly measurable at the Earth.
In fact Eq.~(\ref{eq:ffa}) implies that
the spectra
$\phi_1(E) = \phi[E, V(t_1)]$
and $\phi_2(E) = \phi[E, V(t_2)]$
observed at times $t_1$ and $t_2$ are related to each other by:
\begin{equation}
\phi_1(E) = \frac{p_1^2}{p_2^2} ~\phi_2[E + |q| \, \Delta V(t_1,t_2)] 
\label{eq:ffa1}
\end{equation}
where $\Delta V(t_,t_2) = V(t_1)-V(t_2)$ is the difference between the
modulation potentials at times $t_1$ and $t_2$,
and $p_1$ and $p_2$ are the momenta
that correspond to the energies $E$ and $E + \Delta V$.
The important point of Eq.~(\ref{eq:ffa1}) is that
the two functions that enter the equality are
directly measurable, and this allows to test the validity
of the model without a knowledge of the LIS spectrum.

It is instructive to consider the ideal case of a LIS spectrum that
is a simple power law in rigidity: $J_0 (\rig) = K\, \rig^{-\alpha}$.
The modulated spectrum of a massless particle takes then the form:
\begin{equation}
\flur (\rig, V) = K \, \rig^2 \; (\rig + |Z| \, V)^{-(2+\alpha)}
\label{eq:flux_mass0}
\end{equation}
(with $Z = q/e$).
This flux grows quadratically in $\rig$ for low
rigidities, reaches a maximum at $\rig^* = 2 \, |Z| \, V /\alpha$, and
for large rigidities becomes asymptotically
a simple power law with constant spectral index $\alpha$.
For a particle with mass $m$ the modulated flux takes the form:
\begin{equation}
 \flur (\rig, V) = K \;|q|^{\alpha+3} \; \frac{\rig^3}{E+ m} \;
 (E + m + |q| \, V) \,
 (E+ |q| \, V)^{-(\alpha+3)/2}
\, (E+ 2 m +|q| \, V)^{-(\alpha+3)/2}
\label{eq:flux_rig}
\end{equation}
(where $E = \sqrt{( q \rig)^2 + m^2} - m$ is the kinetic energy that corresponds
to rigidity $\rig$). This form has shape similar to the massless case,
with a flux that grows rapidly for small $\rig$,
reaches a maximum (at a rigidity that grows with $V$)
and then, for large $\rig$, becomes a power law of spectral index $\alpha$.

Expressing the spectrum in terms of kinetic energy, it takes the form:
\begin{equation}
 \phi(E,V) = K \; |q|^{\alpha -1} \;
 E\; (E + 2 m ) \,
 (E + m +|q| \, V) \,
 (E + |q| \, V )^{-(\alpha+3)/2}
\, (E+ 2 m +|q| \, V)^{-(\alpha+3)/2} ~.
\label{eq:flux_e}
\end{equation}

It should be stressed that the expressions (\ref{eq:flux_rig}) and~(\ref{eq:flux_e})
for a rigidity or kinetic energy spectrum might appear as rather complicated, but
they describe a very simple model: an exact power law in rigidity (with normalization $K$ and
spectral index $\alpha$) modulated by a constant energy loss $|q| \, V$.
Adopting these expressions to fit the time dependent rigidity spectra
measured by AMS02 and PAMELA is surprisingly successful.

Fitting the 2717 daily proton and helium daily spectra
with data in the rigidity ranges [1--100]~GV for $p$ (30 bins), and 
[1.71--100]~GV for helium (26 bins) with the form (\ref{eq:flux_rig})
and allowing all three parameters ($K$, $\alpha$ and $V$)
to be time dependent, one obtains reasonably good fits
with global $\chi^2_{\rm min}/{\rm d.o.f.} = 0.86$ for protons
and 0.79 for helium nuclei.
In the case of helium one has the problem that the flux is
formed by a mixture of the two isotopes ${}^4$He
and ${}^3$He \cite{ams_helium_isotopes}.
In this paper we have neglected the rigidity dependence of
the isotopic composition, and assumed a constant ratio
${}^3$He/${}^4{\rm He} \simeq 0.2$.

For the electron daily spectra data, the AMS02 collaboration has
released 3193 spectra in the rigidity range [1--42]~GV. Selecting the smaller
rigidity range $\rig < 10$~GV,
the data can be successfully fitted with the expression
(\ref{eq:flux_rig}) obtaining a global $\chi^2_{\rm min}/{\rm d.o.f.} = 0.68$.
In this case the range of the fit must be reduced because the $e^-$ spectrum has a hardening
that begins at $\rig \simeq 10$~GV \cite{ams_electrons}.

Fitting the CR spectra with the form (\ref{eq:flux_rig}) and three time
dependent parameters can be useful, but it is not entirely satisfactory, because
it is not obvious how to interpret the time dependence of the three parameters
$K$, $\alpha$ and $V$. If one tries to test a ``minimal model'' based on
the FFA model, with $K$ and $\alpha$ constant and a time dependent
(but constant in rigidity) potential one obtains fits that describe
the data reasonably well, with deviations of order 10\%, however, because
of the remarkable accuracy of the AMS02 and PAMELA measurement (with errors
of order 1--3\%), the quality of the fits are poor.

This suggests to introduce a simple generalization of the FFA model, 
that is always based on expression Eq.~(\ref{eq:flux_rig}) to
fit the rigidity spectra, but keeping $K$ and $\alpha$ as time independent
(because they are considered as parameters associated to the LIS spectra)
and introducing a rigidity dependence for the potential $V(t)$.
For this purpose we introduce the form:
\begin{equation}
V(\rig, t) = V_0 (t) + [V_\infty (t) - V_0(t)] ~(1 -e^{-\rig/\rig^*} )
\label{eq:vrig}
\end{equation}
that contains two time dependent parameters $V_0(t)$ and $V_\infty (t)$
that can be interpreted as the average energy losses  (divided by $|q|$)
during propagation in the heliosphere for particles that arrive at the Earth with
very small and very large rigidities.
It is also possible to express the potential in
terms of $V_1 = V(\rig_1)$ and $V_2 = V(\rig_2)$, that are the values of $V$
for two (arbitrary, but conveniently chosen) rigidities: 
\begin{equation}
 V(\rig, t) = \frac{e^{-\rig/\rig^*}}{e^{\rig_1 / \rig^*} -e^{\rig_2 / \rig^*}}
 ~\left [
 V_1(t) \;\left (
 e^{(\rig +\rig_1 )/\rig^*} -e^{(\rig_1 + \rig_2)/\rig^*} \right )
 -V_2(t) \;\left (
 e^{(\rig +\rig_2 )/\rig^*} -e^{(\rig_1 + \rig_2)/\rig^*} \right )
 \right ] 
\label{eq:vrig1}
\end{equation}
The AMS02 data are published for rigidities larger than 1~GV, and
the effect of modulations are small and difficult to measure for $\rig \gg 10$~GV,
and therefore in the present paper, we have chosen to parametrize the energy dependence of
the potential with $V_1 = V(1~{\rm GV})$ and $V_2 = V(10~{\rm GV})$.
The potential in Eq.~(\ref{eq:vrig1}) also contains the additional parameter
$\rig^*$, that is kept constant with value $\rig^* = 6$~GV \footnote{Considering
 $\rig^*$ as a free parameter improve significantly the quality
 of the fits only for a small number of the spectra in the AMS02 data set.}.

In the remaining of this paper we will fit the lower rigidity
part of the CR spectra for protons, helium nuclei electrons and positrons
with the scheme we have outlined, that is using 
Eq.~(\ref{eq:flux_rig}) with a 
a time dependent potential of form~(\ref{eq:vrig1}).
For the two parameters $K$ and $\alpha$ that describe the
power law spectra, we have used the average values $\langle K\rangle $
and $\langle \alpha \rangle$ obtained from fits
to all AMS02 spectra
based on Eq.~(\ref{eq:flux_rig})
with all three parameters $K$, $\alpha$ and $V$ free (and $V$ constant in rigidity).
The results are:
$K = 2.94$, 0.426, 0.743 and $5.01\times 10^{-3}$ [in units (cm$^{2}$ \,s \,sr\, GV)$^{-1}$]
and 
$\alpha = 2.90$, 2.80, 4.10 and 3.42 
for $p$, He, $e^-$ and $e^+$ respectively.

With this scheme one obtains reasonably good fits to all AMS02 and PAMELA observations.
For example, the global chi squared of fits to the AMS02 daily spectra are
$\chi^2_{\rm min}/{\rm d.o.f.} = 0.82$, 0.76 and 0.71 for $p$, He and $e^-$.
These values are approximately equal to those obtained using
Eq.~(\ref{eq:flux_rig}) with three time dependent parameters:
$K$, $\alpha$ and (constant in rigidity) $V$, but the
interpretation of the parameters is now more natural.

\begin{figure}
\begin{center}
\includegraphics[width=14cm]{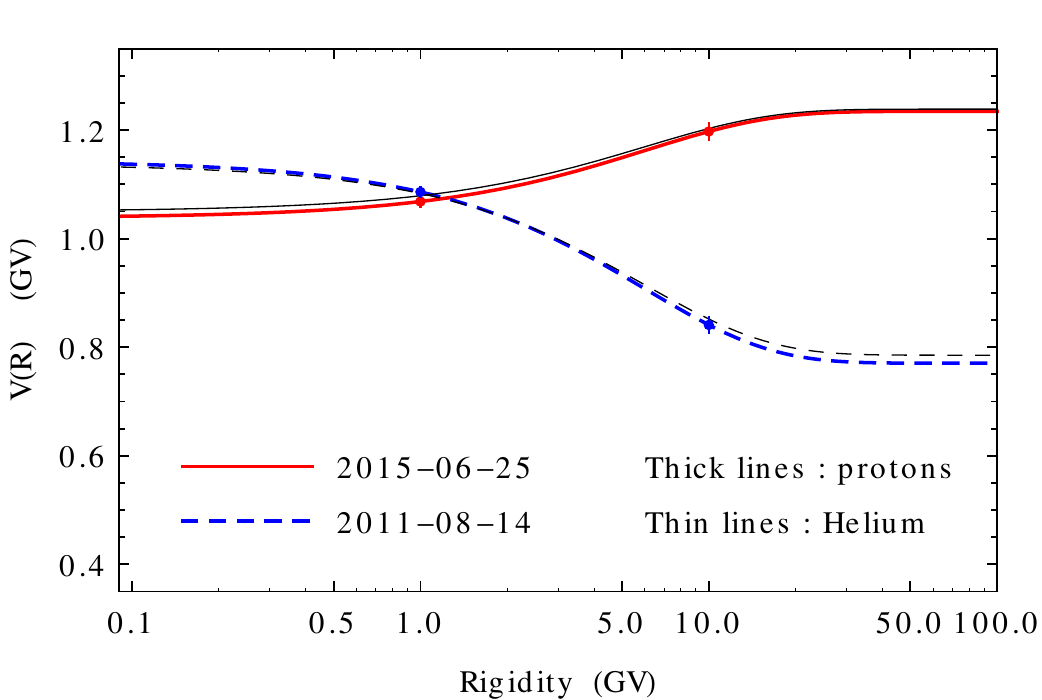}
\end{center}
\caption {\footnotesize
 Potentials obtained fitting the proton and helium spectra
 shown in Fig.~\ref{fig:flux_example} with expression
 of Eq.~(\ref{eq:flux_rig}).
 The rigidity dependence of the potentials has the form
 of Eq.~(\ref{eq:vrig1}).
 \label{fig:v_example}}
\end{figure}

Four examples of fits to the AMS02 measurements (two for $p$ spectra, and
two for He spectra)
are shown in Fig.~\ref{fig:flux_example},
where one can see that they give a good description of the data.
The rigidity dependent potentials with form
(\ref{eq:vrig1}) that enter the expression for the rigidity spectrum
of Eq.~(\ref{eq:flux_rig}) are shown in Fig.~\ref{fig:v_example},
where the points show the best fit values of the parameters $V_1$ and $V_2$.
One can see that the potentials have a modest but significant 
rigidity dependence with a form that is different for
spectra observed at different times.
It is remarkable that the potentials 
obtained fitting the $p$ and He spectra measured the same day are (within errors)
equal to each other. This is in fact a result that is in general valid for all the $p$
and He daily spectra measured by AMS02, indicating that the solar modulations
for protons and helium nuclei are in good approximation equal.
It should be noted that the potential that enters the expression
of Eq.~(\ref{eq:flux_rig}) for the modulations is multiplied by the 
absolute value of the electric charge of the particles,
therefore this result can also be stated saying that
(in an appropriate sense) the effects of solar modulations are two times larger 
for helium (that has charge number $Z=2$) .

Some other examples of fits to AMS02 and PAMELA spectra
for protons, helium nuclei, electrons and positrons 
calculated in the scheme we are discussing here 
are shown in Fig.~\ref{fig:voyager}.
In the figure the spectra and their fits are shown, as function of kinetic energy, in four
separated panels, where the power law rigidity spectra that enters
the expression of Eq.~(\ref{eq:flux_e}) are also shown as dashed lines.
In three panels (for $p$, He and $e^-$) 
we also show the measurements obtained by Voyager~1
after crossing the heliopause at a distance of approximately 120~AU from the Sun
\cite{voyager-2016} that are considered as representative of the CR spectra in the
local interstellar medium. 

Each one of the panels include three spectra from AMS02.
For $p$, He and $e^-$ the three spectra are 
the highest, the lowest and an intermediate one, chosen among the daily measurements
\cite{ams_daily_protons,ams_daily_helium,ams_daily_electrons}.
For positrons, the three spectra are again the highest, lowest and an intermediate one, but
chosen among the measurements obtained averaging over
one Bartels rotation \cite{ams_bartels_electrons}.
In the panels for $p$ and $e^-$ we include two spectra
(the highest and lowest) obtained by PAMELA \cite{Adriani:2013as,Martucci:2018pau,Adriani:2015kxa}
with longer averaging times.
The PAMELA results are of great interest because they cover
a different time interval (June--2006 to January--2018)
and because they are available in a kinematic range
that extends to lower rigidities.
Our model gives a good description also of the lower rigidity
observations of PAMELA, with significant deviations
only for the electron spectra at $E \lesssim 200$~MeV.

\section{Time dependence of the Potentials}
\label{sec:potentials}
\begin{figure}
\begin{center}
 \includegraphics[width=7.0cm]{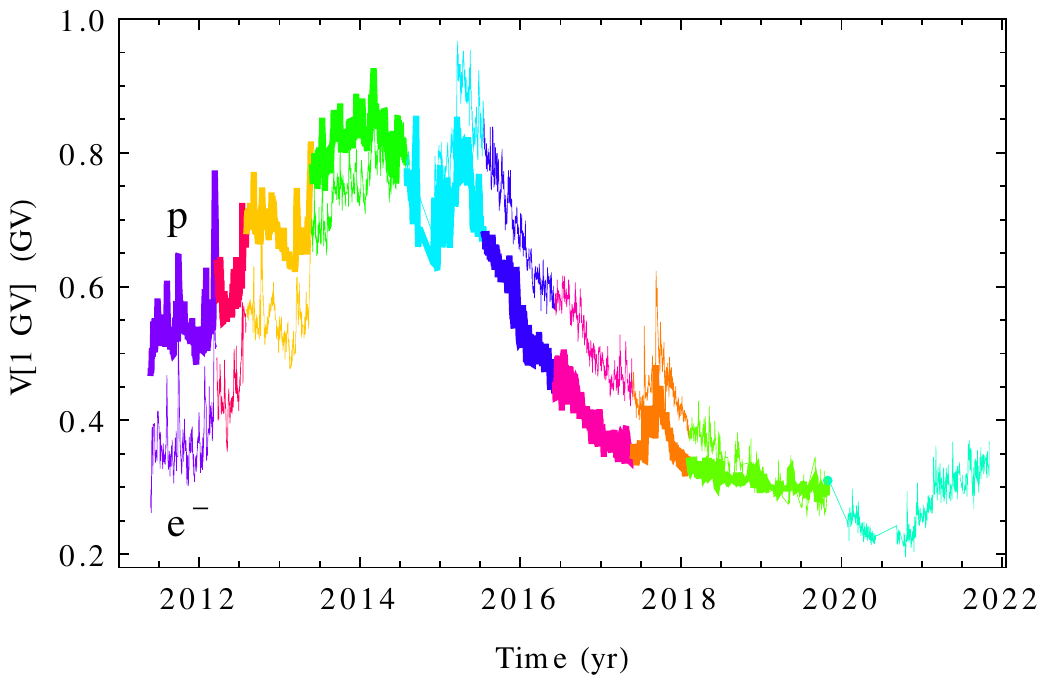}
 ~~~
\includegraphics[width=7.0cm]{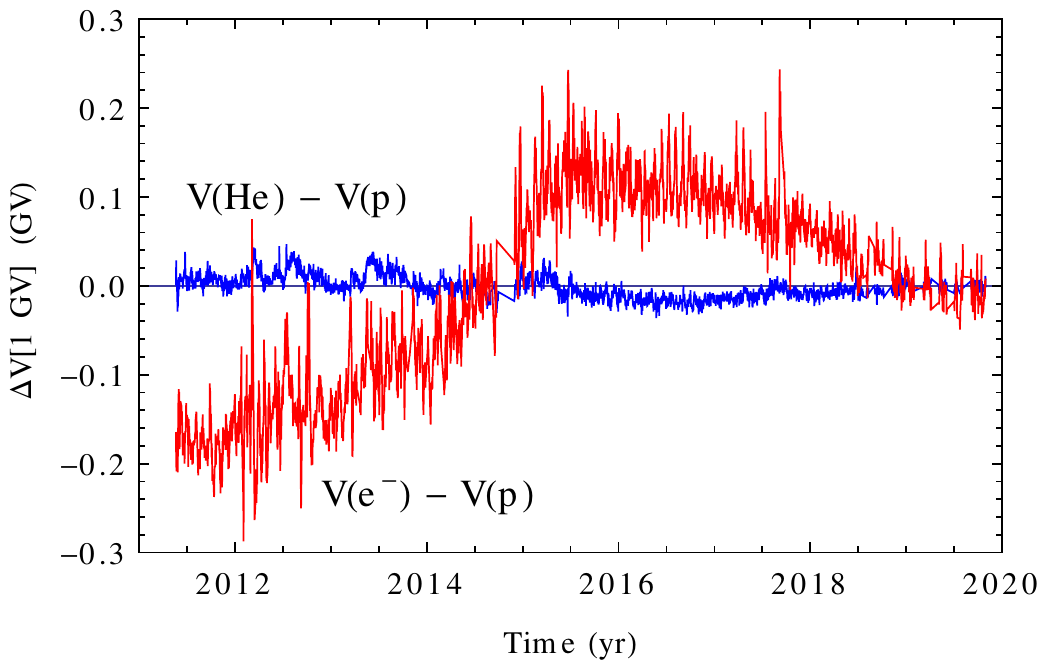}
\vspace{0.5 cm}
\includegraphics[width=7.0cm]{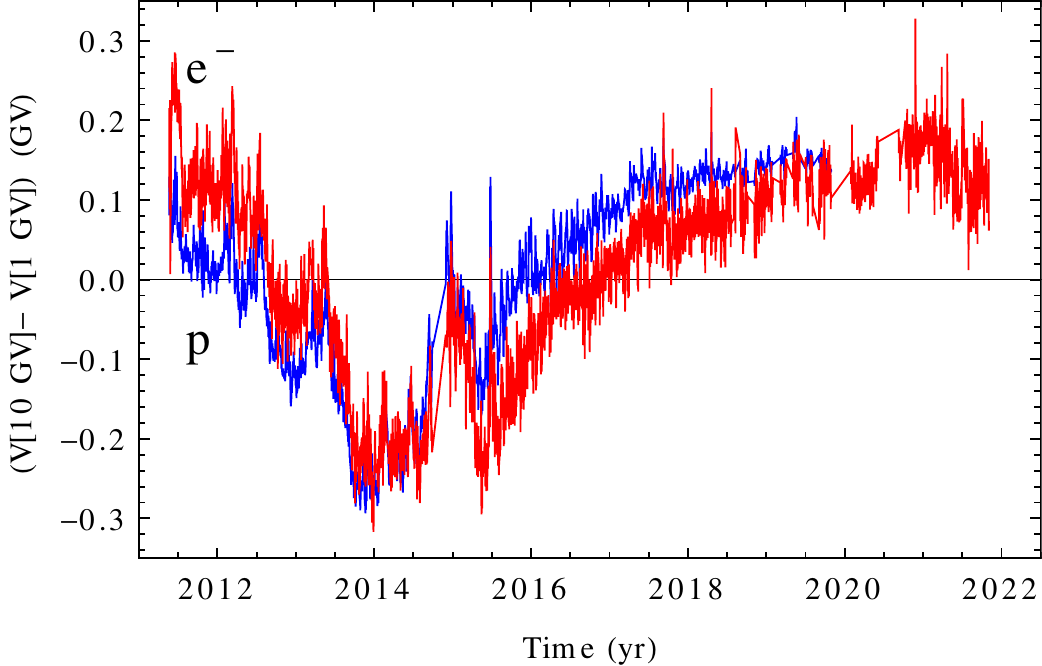}
~~~
\includegraphics[width=7.0cm]{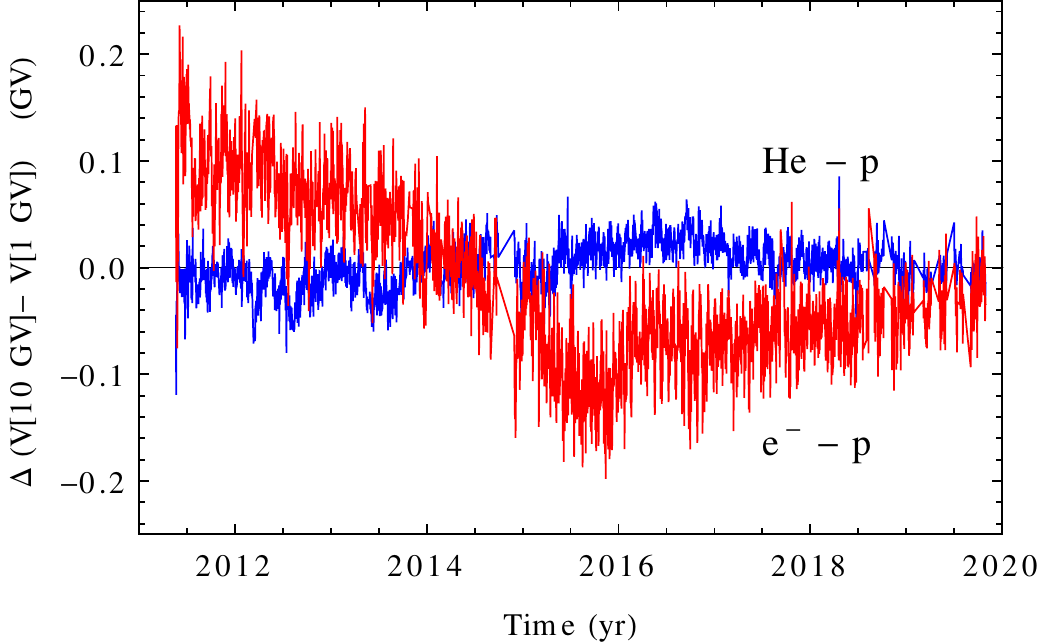}
\end{center}
\caption {\footnotesize
 Parameters of the potentials obtained from fits to
 the AMS02 measurements of the daily spectra of
 protons, helium nuclei and electrons
\cite{ams_daily_protons,ams_daily_helium,ams_daily_electrons}.
The top--left panel shows the potential $V_{[1~{\rm GV}]}$
(thick line for $p$, thin line for $e^-$, 
 the colors identify different characteristic time intervals).
 The top--right panel shows the differences
 $(V_{[1~{\rm GV}]}^{\rm He} - V_{[1~{\rm GV}]}^{\rm p})$ 
 and
 $V_{[1~{\rm GV}]}^{\rm e^-} - V_{[1~{\rm GV}]}^{\rm p}$. 
The bottom--left panel shows the potential difference
$\Delta V = V_{[10~{\rm GV}]} - V_{[1~{\rm GV}]}$ for
protons and elecrons.
The bottom--right panel shows the differences
$(\Delta V^{\rm He} - \Delta V^{p})$ and 
$(\Delta V^{\rm e^-} - \Delta V^{p})$.
 \label{fig:v_daily}}
\end{figure}

The time dependence of the potentials obtained fitting the daily spectra measured by
AMS02 for protons, helium nuclei and electrons are shown in Fig.~\ref{fig:v_daily}.
The potentials in the figure include the subtraction of a constant shift 
that depends on the particle type:
($\Delta V_{\rm LIS} =0.29$, 0.30 and 1.14~GV for $p$, He and $e^-$ respectively)
that will be discussed in the next section.

The top--left panel in Fig.~\ref{fig:v_daily} shows the
potential $V_1 = V_{[1~{\rm GV}]}$ for protons and electrons.
The two potentials have significantly different time dependences,
and with the shifts that we have introduced
are approximately equal during the time interval,
in the middle of 2014, that corresponds
to the reversal of the polarity of the solar magnetic field.
One also has
(for both rigidities 1~GV and 10~GV) the inequalities:
\begin{equation}
 \begin{cases}
 V^{(e^-)} (t) < V^{(p)} (t) & \text{for} ~~ t < t_{\rm reversal} \\[0.12cm]
 V^{(e^-)} (t) > V^{(p)} (t) & \text{for} ~~ t > t_{\rm reversal} 
 \end{cases}
\end{equation}
At the reversal time $t_{\rm reversal}$
the solar magnetic field polarity changes from negative ($A = -1$) to
positive ($A = +1$).
During a phase of negative polarity particles with electric charge
$q < 0$ arrive at the Earth from the heliospheric poles, while
particles with $q > 0$ arrive travelling close to the heliospheric equator
and the wavy current sheet.
The situation is reversed after the flip of the magnetic field polarity.
Our results are therefore consistent with the expectation that the energy losses
during propagation in the heliosphere are larger for particles
that arrive from the heliospheric equator \cite{Potgieter:2013pdj,Lipari:2014gfa}.

The top--right panel in Fig.~\ref{fig:v_daily} shows the
differences between the potentials at rigidity 1~GV
of electrons and protons and of helium nuclei and protons.
It is striking that the potentials of $p$ and He are approximately equal.
This result has important implications, because it validates the
idea of using a potential to describe solar modulations, and
is consistent with models where protons and helium nuclei
of equal rigidities follow (approximately) equal trajectories
in the heliosphere.

The bottom--left panel in Fig.~\ref{fig:v_daily} shows the time
dependence of the potential differences $\Delta V = V_{[10~{\rm GV}]} - V_{[1~{\rm GV}]}$
for protons and electrons.
The rigidity dependence of the potentials is rather small
(with $|\Delta V | \lesssim 0.25$~GV), so that a simple FFA parametrization can be considered,
for many applications, a reasonable approximation, validating many studies performed in the past,
however the introduction of a rigidity dependence is necessary to obtain good quality fits.
It is also important to note that $\Delta V$ can be either positive or negative at different times,
so that the modulated spectra can have different shapes at different times.

The bottom--right panel in Fig.~\ref{fig:v_daily} shows the difference
between $\Delta V$ for electrons and protons, and for helium nuclei and protons.
One can note that the rigidity dependences of the potentials for electrons
and protons are strongly correlated but not identical. This can be understood
as the consequence of the facts that in general the properties of the
(different) regions of the heliosphere where particles of opposite electric
charge propagate are correlated, for example because the same CME's
can perturb both regions.
The difference in $\Delta V$ between protons and helium nuclei is much smaller,
and again indicates that the solar modulation effects are in good approximation
equal for the two particles.

\begin{figure}
\begin{center}
\includegraphics[width=7.5cm]{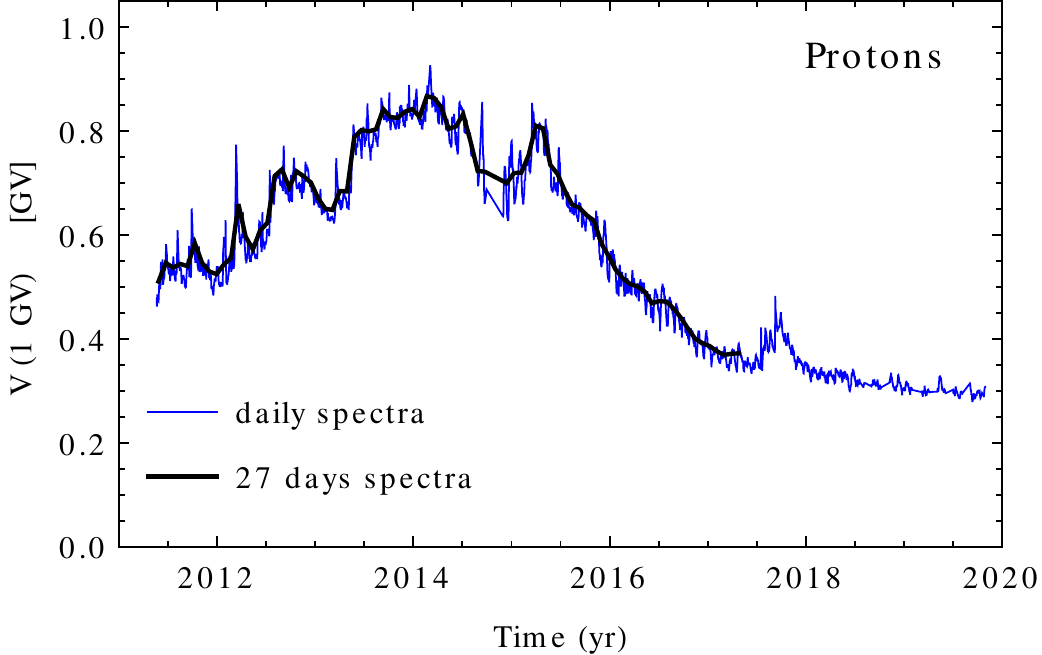}
~~~~
\includegraphics[width=7.5cm]{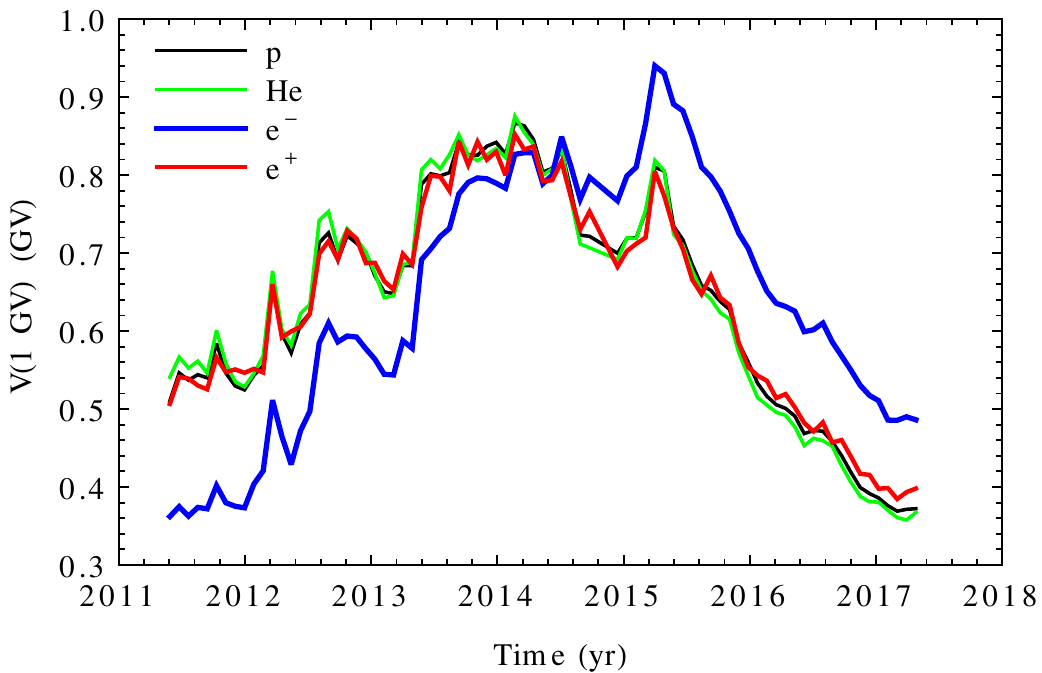}

\vspace{0.8 cm}
\includegraphics[width=7.5cm]{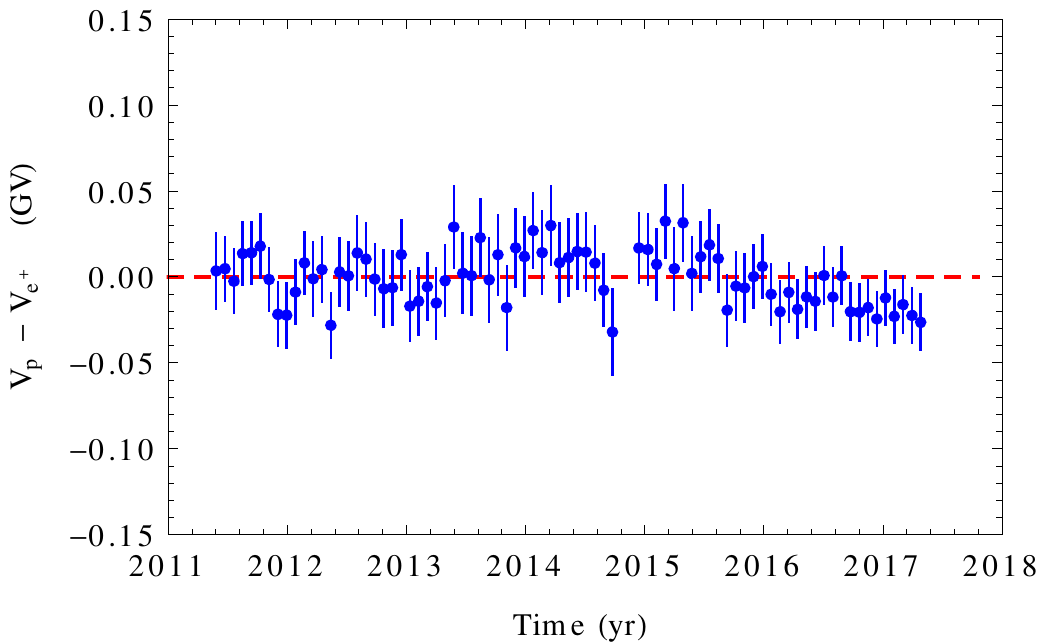}
~~~
\includegraphics[width=7.5cm]{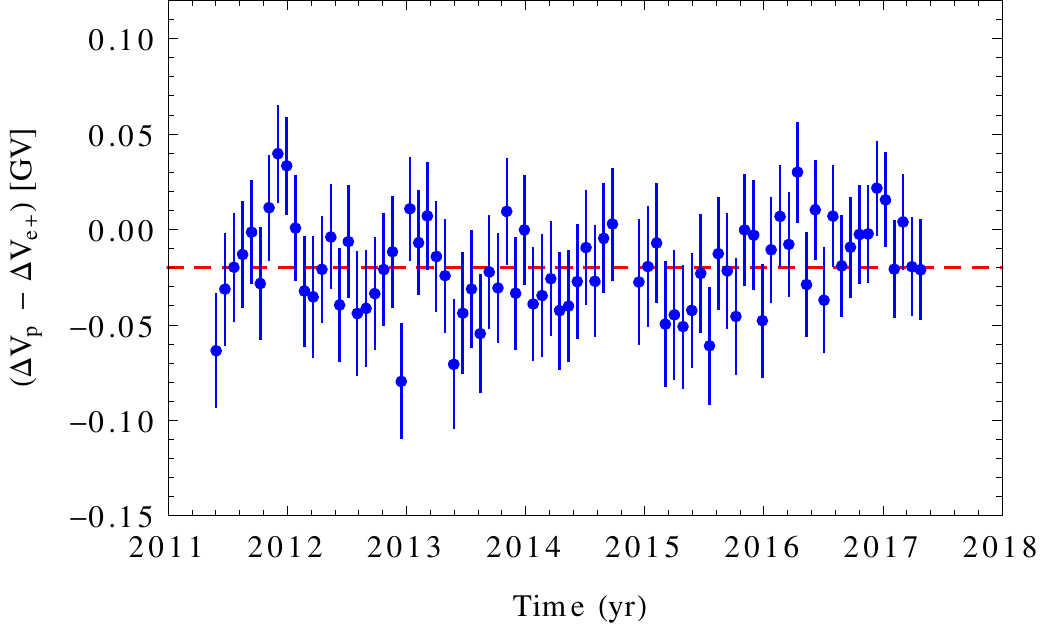}
\end{center}
\caption {\footnotesize
 The top--left panel shows the potential $V_{[1~{\rm GV}]}$
 for protons calculated fitting the 27 days averaged
 spectra \cite{ams_bartels_protons} and the daily averaged spectra
 \cite{ams_daily_protons} measured by AMS02.
 The top--right panel compares the potentials 
 $V_{[1~{\rm GV}]}$ obtained fitting the 
 27 days averaged spectra for $p$, He, $e^-$ and $e^+$
 \cite{ams_bartels_protons,ams_bartels_electrons}
 (error bars are not shown).
 The two panels at the bottom compare the proton and
 positron potentials.
 The bottom--left panel shows the difference
 $(V_p - V_e^+)$ at rigidity 1~GV;
the bottom--right panels show the difference
 and $(\Delta V_{p} - \Delta V^{e^+})$ 
 (with $\Delta V = (V_{[1~{\rm GV}]} - V_{[10~{\rm GV}]})$.
 \label{fig:cycles}}
\end{figure}

To study positron solar modulation we have fitted
the AMS02 measurements of $p$, He, $e^-$ and $e^+$ spectra
obtained averaging over 27 days \cite{ams_bartels_protons,ams_bartels_electrons}.
The results are shown in Fig.~\ref{fig:cycles}.
In the top--left panel the proton potential at rigidity $\rig \simeq 1$~GV
is compared to the one obtained fitting the daily spectra
to show the consistency of the results.
In the top--right panel the potentials (always at 1~GV) for
the four particles ($p$, He, $e^-$ and $e^+$) are shown together,
with the potential for positrons shifted by $\Delta V_{\rm LIS}^{(e^+)} \simeq 0.176$~GV).
The potentials for the three positively charged particles
($p$, He and $e^+$) are in good approximation equal,
while the potential for $e^-$ is significantly different.

It should be noted that one expects that the modulations of particles
with the same electric charge but different mass cannot be identical,
with differences that increase in importance for low rigidities.
The differences in modulation are expected because the relation between energy
and rigidity is mass dependent, so that particles of different
mass that enter the heliosphere at the same point with the same initial $\rig_i$
will develop different rigidities due to energy losses, and travel along different trajectories.
In addition, particles with identical rigidity but different mass will have
different velocities and therefore different propagation times in the heliosphere,
and this can also result in different
modulations if the heliosphere is not in a stationary state.
Our analysis shows only small differences in the potentials for protons
and helium (at the level of a few percents).
Future studies of these mass dependent effects that include helium nuclei
will also have to take into account their rigidity dependent isotopic composition.

\begin{figure}
\begin{center}
\includegraphics[width=12.0cm]{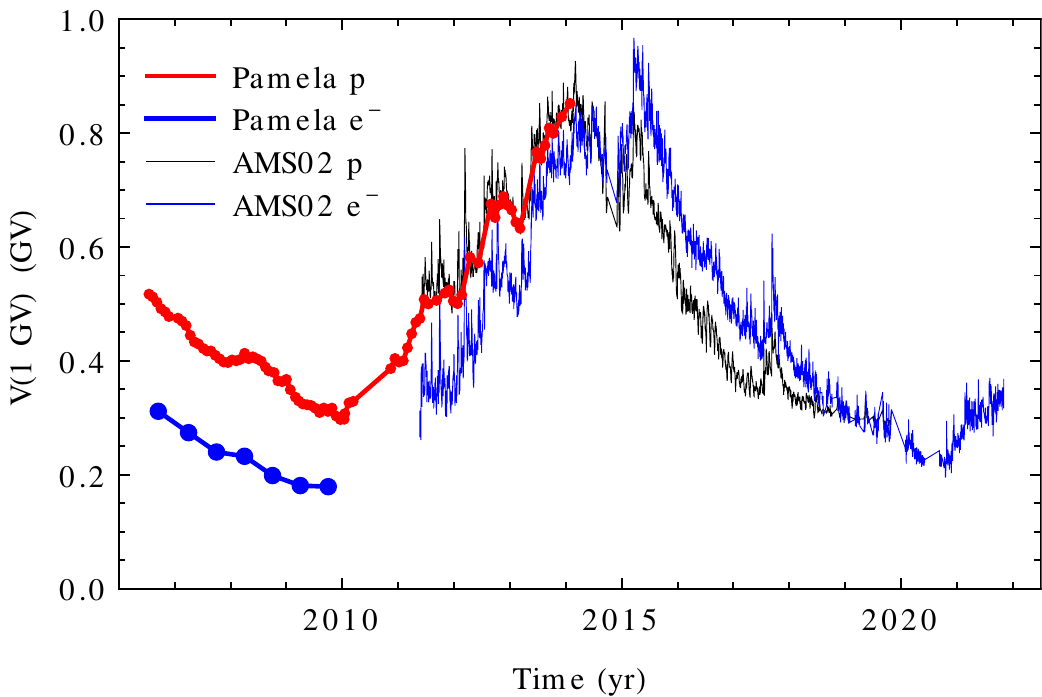}
\end{center}
\caption {\footnotesize
 Plot of the time dependence of the potential at rigidity 1~GV obtained from fits to
 the spectra of protons and electrons.
 The data are are from PAMELA \cite{Adriani:2013as,Martucci:2018pau,Adriani:2015kxa,Adriani:2016uhu}
 and AMS02~\cite{ams_daily_protons,ams_daily_electrons}.
\label{fig:v_pam_ams} }
\end{figure}

The results of the potentials at 1~GV for fits to the PAMELA protons (83 spectra
\cite{Adriani:2013as,Martucci:2018pau})
and electrons (7 spectra \cite{Adriani:2015kxa})
are shown, together with the fits to the AMS02 daily spectra, 
in Fig.~\ref{fig:v_pam_ams}.
The PAMELA data start in June 2006, and
cover also the final part of solar cycle 23. The measurements of the
proton spectra extend to the beginning of 2014, and can be 
compared with the first part of the AMS02 data.
The agreement between the two data sets is good.
The measurements of the $e^-$ spectrum extend only to 2009, and
such a comparison is not possible.

\section{The Local Interstellar Spectra}
\label{sec:lis}

It is now desirable, indeed necessary, to address the question of what physical
meaning can be attributed to the potentials we have obtained
fitting the AMS02 and PAMELA data, and what 
can be deduced from these studies about the CR interstellar spectra.

In the FFA model the physical meaning of the (rigidity independent
in the original formulation) potential is clear: it gives the
average energy loss (divided by $|q|$) suffered by CR particles
in their propagation from the boundary of the heliosphere to the Earth.
In the model discussed here the potential describes a
spectral distortion calculated with respect to an
``artificial'' spectrum, that has a simple power law form in rigidity,
and therefore this potentials does {\em not} have a well defined physical meaning.
However, the difference $\Delta V(\rig, t_1,t_2) = V(\rig, t_1) - V(\rig, t_2)$ between
potentials obtained from fits to the spectra measured at times $t_1$ and $t_2$,
is related to the two observed spectra via Eq.~(\ref{eq:ffa1}),
and can be interpreted as the difference in the average energy loss suffered during
heliospheric propagation by particles
observed with rigidity $\rig$ at times $t_1$ and $t_2$.
The simple power law spectrum ``cancels'' in this comparison,
as it plays the role of a ``scaffolding'', used to perform the fits
and obtain the potentials, and that can then be discarded.

This procedure leaves the LIS spectra undetermined, and this is a
serious limitation because the determination of the interstellar
spectra is a fundamental goal in the study of solar modulations.

There is a large literature about estimating the shape
of the cosmic ray LIS spectra
(see for example \cite{Boschini:2017fxq,Bisschoff:2019lne})
and in all these studies the measurements
obtained by Voyager~1 beyond the heliopause \cite{voyager-2016}
play a crucial role.
One should however note that the Voyager data, while of great value,
are not sufficient to allow a model independent determination of the LIS spectra.
This is because the Voyager data cover only a limited kinematical range
(a maximum observed energy of 350~MeV for protons, and 75~MeV for electrons).
Since the energy lost by CR particles traversing the heliosphere
is of order 300~MeV or more, it follows that the CR particles in
the range observed by Voyager do not reach the Earth, and vice--versa
the particles in the energy range of observations at the Earth 
arrived at the boundary of the heliosphere with energy 
above the range of the Voyager measurements, and therefore a direct
comparison of shapes of the spectra formed by the same particles
in interstellar space and at the Earth is not possible.

\begin{figure}
\begin{center}
\includegraphics[width=7.6cm]{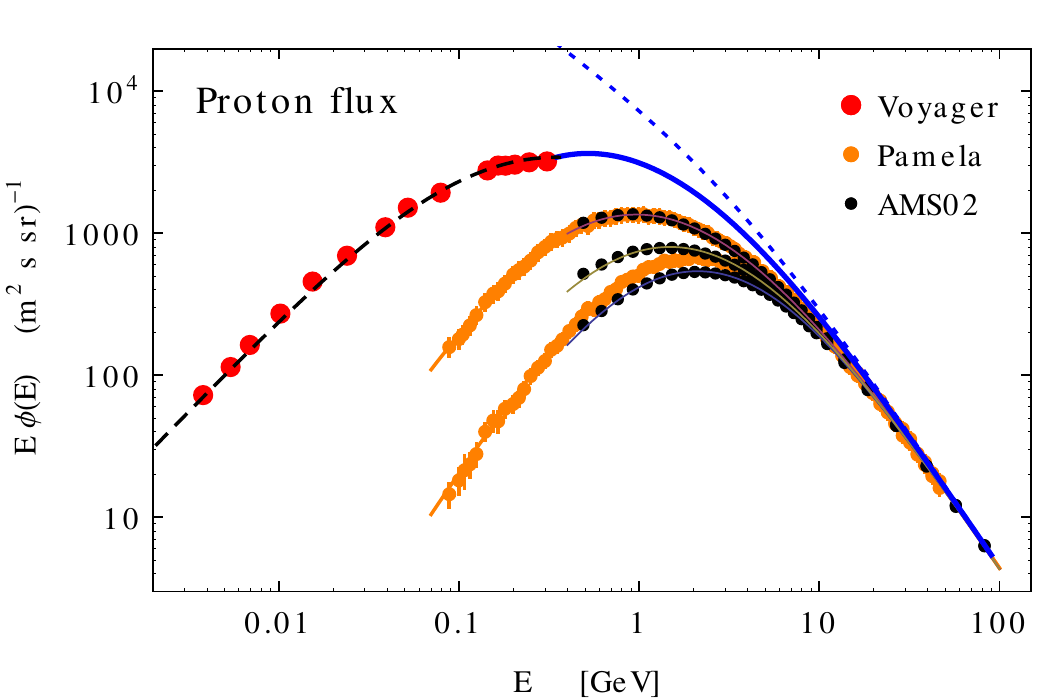}
~~
\includegraphics[width=7.6cm]{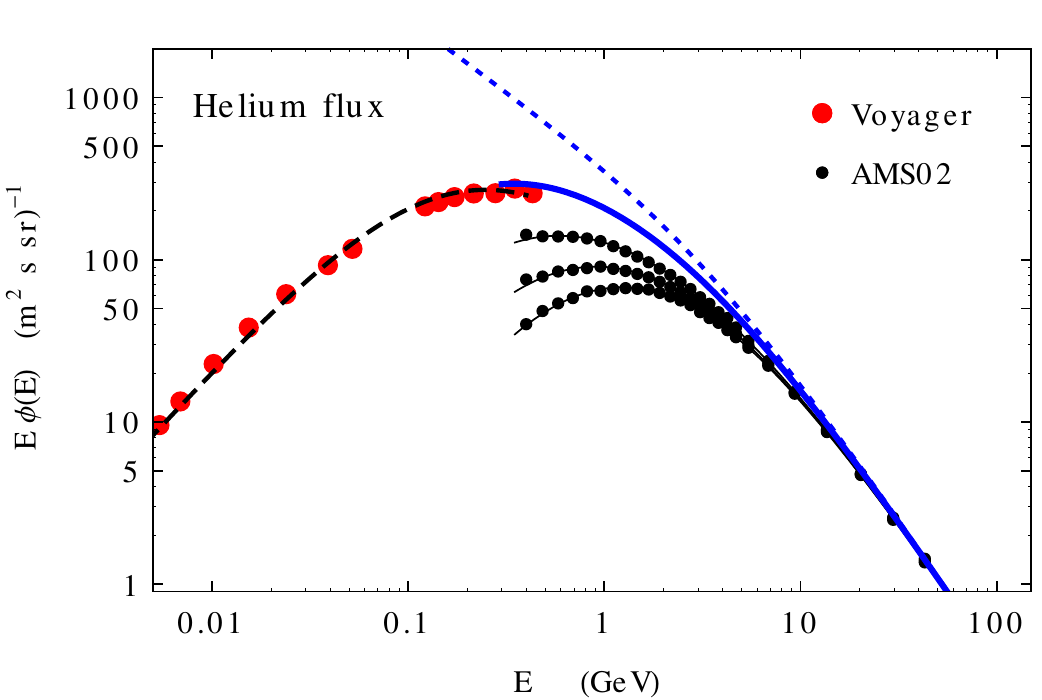}

\vspace{0.9 cm}
\includegraphics[width=7.6cm]{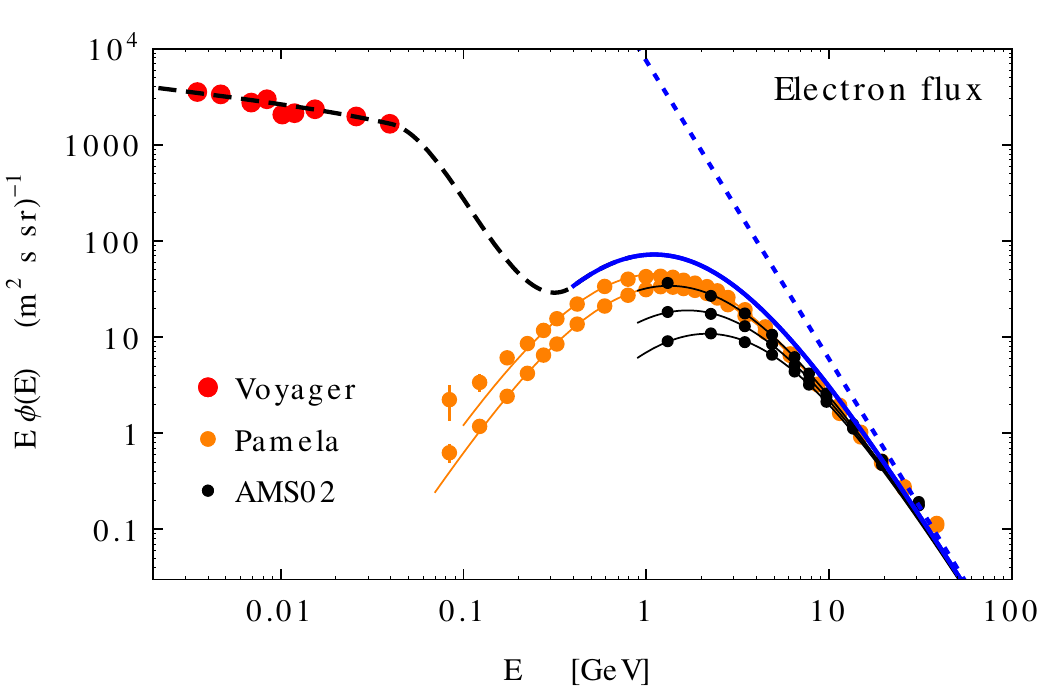}
~~
\includegraphics[width=7.6cm]{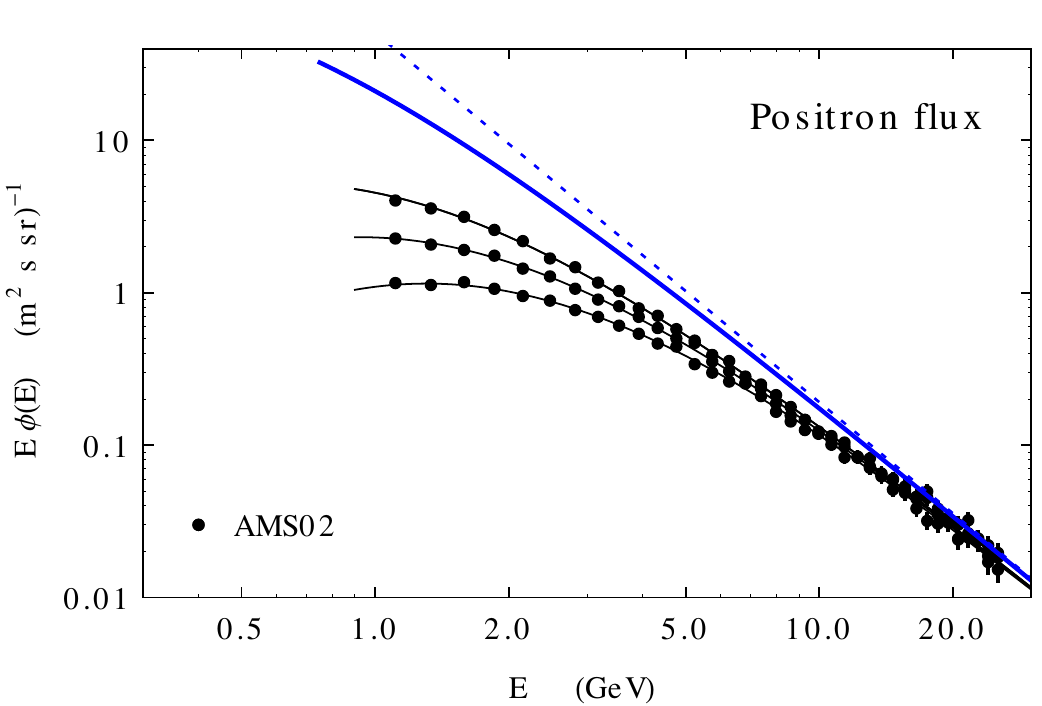}
\end{center}
\caption {\footnotesize Spectra of protons, Helium nuclei, electrons
 and positrons plotted as a function of kinetic energy. The data are from
 Voyager (\cite{voyager-2016} for $p$, He and $e^-$),
 PAMELA (\cite{Adriani:2013as,Martucci:2018pau} for
 $p$ and \cite{Adriani:2015kxa} for $e^-$) and AMS02
 (\cite{ams_daily_protons} for $p$,
 \cite{ams_daily_helium} for He,
 \cite{ams_daily_electrons} for $e^-$ and
 \cite{ams_bartels_electrons} for $e^+$).
 For AMS02 we show the highest and lowest spectrum among those published,
 and a third spectrum that is approximately the geometric average of the
 first two. For PAMELA we show the highest and lowest spectrum.
 The dotted lines show spectra that are simple power laws in rigidity.
 The thin lines are fits to the AMS02 and PAMELA data
 using the form Eq.~(\ref{eq:flux_e}) 
 with the potential of Eq.~(\ref{eq:vrig1}).
 The thick dashed lines  are fits to the Voyager data.
 The thick solid lines are calculated using
 the form of Eq.~(\ref{eq:flux_e}) 
 with a constant potential
chosen to connect smoothly to  the Voyager data
(see discussion in the main text).
 For electrons  the connection requires  the existence of structure in the
 $e^-$ LIS spectrum.
 \label{fig:voyager} }
\end{figure}

The Voyager data are of course a very important constraint
in the construction of the LIS spectra. The importance of this
constraint is evident comparing the spectra in Fig.~\ref{fig:voyager}.
For example inspecting the top--left panel in the figure
one can see that the proton rigidity power law spectrum
(shown as a dotted line) used as a starting
point in the fitting procedure, is clearly much larger that the LIS
spectrum. On the other hand, distorting this power law spectrum
with a rigidity independent potential of 0.29~GV one obtains
(thick solid line) a spectrum that joins smoothly the Voyager data.

The same considerations are valid for the helium spectrum,
where distorting the power law spectrum with a rigidity independent potential of
0.30~GV one obtains a flux that joins smoothly the Voyager data
(see the top--right panel in Fig.~\ref{fig:voyager}).

This suggests that the LIS spectra for protons and helium
can be, in first approximation, described by a power law in rigidity
distorted by a rigidity independent potential $\Delta V_{\rm Lis}$.
The potential $V_{\rm fit} (\rig, t)$ obtained from a fit connects
the spectrum observed at time $t$ to a simple power law spectrum;
subtracting the shift $\Delta V_{\rm LIS}$ one
obtains a potential $V(\rig,t)$ that connects the observed and the interstellar
spectra, and therefore (in first approximation) describes the energy losses
of the CR particles during heliospheric propagation:
\begin{equation}
V (\rig, t) \simeq V_{\rm fit} (\rig, t) - \Delta V_{\rm LIS} ~.
\end{equation}

Extending these considerations to the electron spectra poses some
very interesting problems. A first consideration is that, as already
discussed, we expect that the potentials for particles
with electric charge of opposite sign will in general be different.
This is because the trajectories of charged particles
are also determined by the regular
heliospheric magnetic field, and particles with opposite electric charge
will propagate in different regions of the heliosphere,
where they can suffer different energy losses.
At solar maximum however, during the reversal of the heliospheric magnetic field
polarity, the regular field is negligible, and the trajectories of
the CR particles are controlled only by the random field.
This implies that during the duration of the polarity reversal,
the potentials for particles of opposite electric charge should be
approximately equal.
Imposing the constraint:
\begin{equation}
\langle V^{(e^-)} \rangle_{\rm reversal} =
\langle V^{(p)} \rangle)_{\rm reversal} 
\end{equation}
for averages of the potentials during the field polarity reversal
(that is approximately the time interval from May to July 2014),
we arrive to an estimate of the potential shift required for electrons:
$\Delta V_{\rm LIS}^{(e^-)} \simeq 1.14$~GV.

The electron LIS spectrum calculated with this shift 
is shown in the bottom--left panel of Fig.~\ref{fig:voyager}.
To connect this estimate of the LIS spectrum to the
Voyager data (that are available only at very low energy: $E \lesssim 75$~MeV)
seems to require a non trivial spectral shape,
perhaps indicating the presence of an additional low energy
component in the electron spectrum.

For positrons no measurements at large distance from the Sun are available
to constrain the shape of the $e^+$ LIS spectrum, however it is possible
to estimate the shift $\Delta V_{\rm LIS}^{(e^+)}$ comparing fits to the
$p$ and $e^+$ spectra taken simultaneously and averaged over
one Bartels rotation \cite{ams_bartels_protons,ams_bartels_electrons}.
The potentials for $p$ and $e^+$ are shown in Fig.~\ref{fig:cycles},
and are consistent with a constant difference:
\begin{equation}
V_{\rm fit}^{(e^+)} (\rig, t) \simeq V_{\rm fit}^{(p)} (\rig, t) - 0.124~{\rm GV}~.
\end{equation}
suggesting that the $\Delta V_{\rm LIS}^{(e^+)} \simeq \Delta V_{\rm LIS}^{(p)} - 0.124$~GV.
Adopting this shift one obtains for positrons the LIS spectrum
shown with the thick solid line in the bottom--right panel in Fig.~\ref{fig:voyager}.

The estimates of the LIS spectra obtained in this section are only tentative,
and are not justified by a theoretical model, and therefore
of limited value. In particular, the assumption that $\Delta V_{\rm LIS}$ is rigidity
independent does not have a good justification,
except for the fact that it results,
for all the four particle types considered here ($p$, helium nuclei, $e^\mp$),
in a remarkably simple form for the LIS spectra,
with a shape determined by only two parameters
(the spectral index $\alpha$ and the potential $\Delta V_{\rm LIS}$).
The possible implications of this result deserve a more detailed study.

The study of the shape of the $e^\mp$ LIS spectra
is of particular importance because different models predict that in the
kinematical range where solar modulations are important
($0.1 \lesssim \rig \lesssim 10$~GV) one should observe
spectral structures, associated for example to the critical energy where
energy losses during interstellar propagation
become the dominant sink mechanism for $e^\mp$
(overtaking escape from the Galaxy) \cite{Lipari:2018usj},
or the critical energy where a new source mechanism
(such as acceleration in Pulsars) becomes the dominant one \cite{DiMauro:2023oqx}.
The simple shape of the $e^\mp$ LIS spectra suggested by our study
disfavours these possibilities.

\section{Hysteresis loops}
\label{sec:hysteresis}

\subsection{The 22--year solar cycle}
\begin{figure}
\begin{center}
  \includegraphics[width=10.0cm]{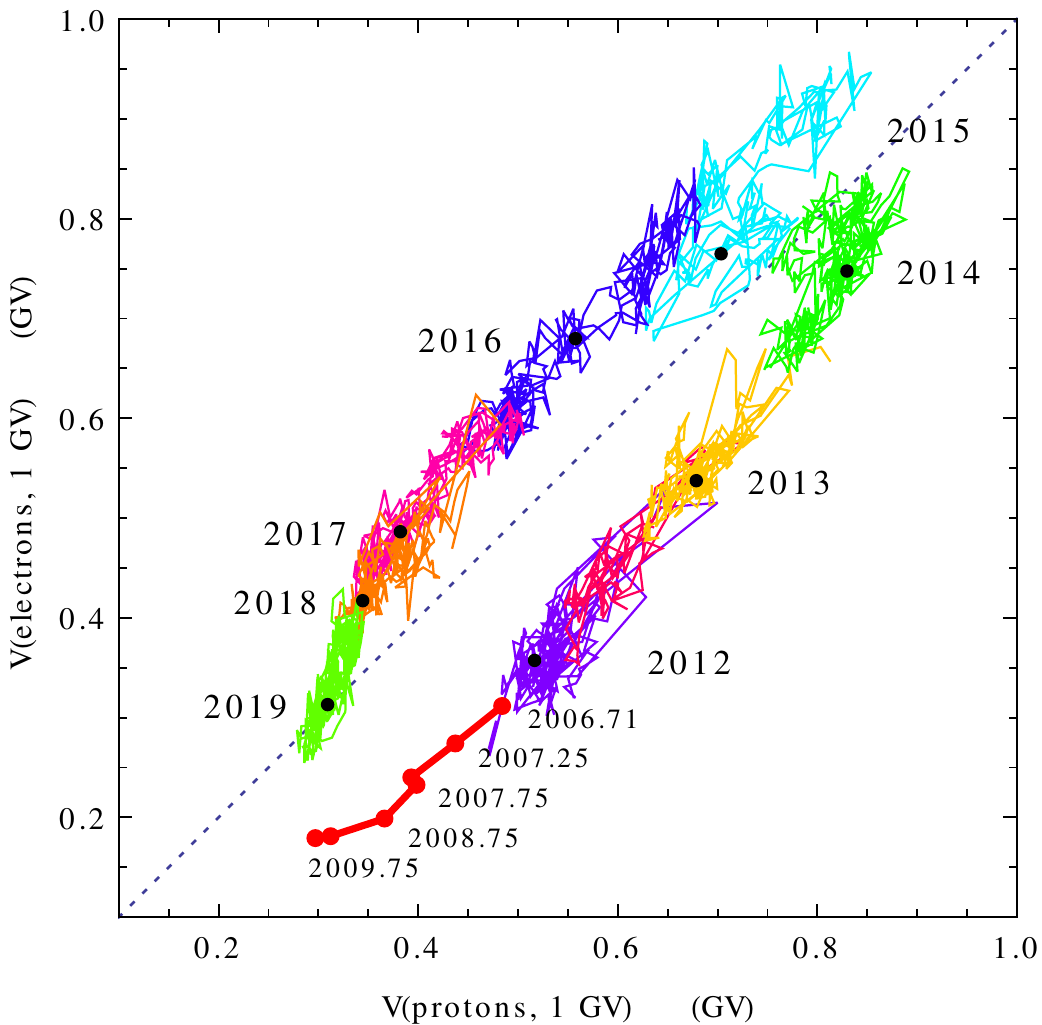}

\vspace{1.0 cm}
\includegraphics[width=10.0cm]{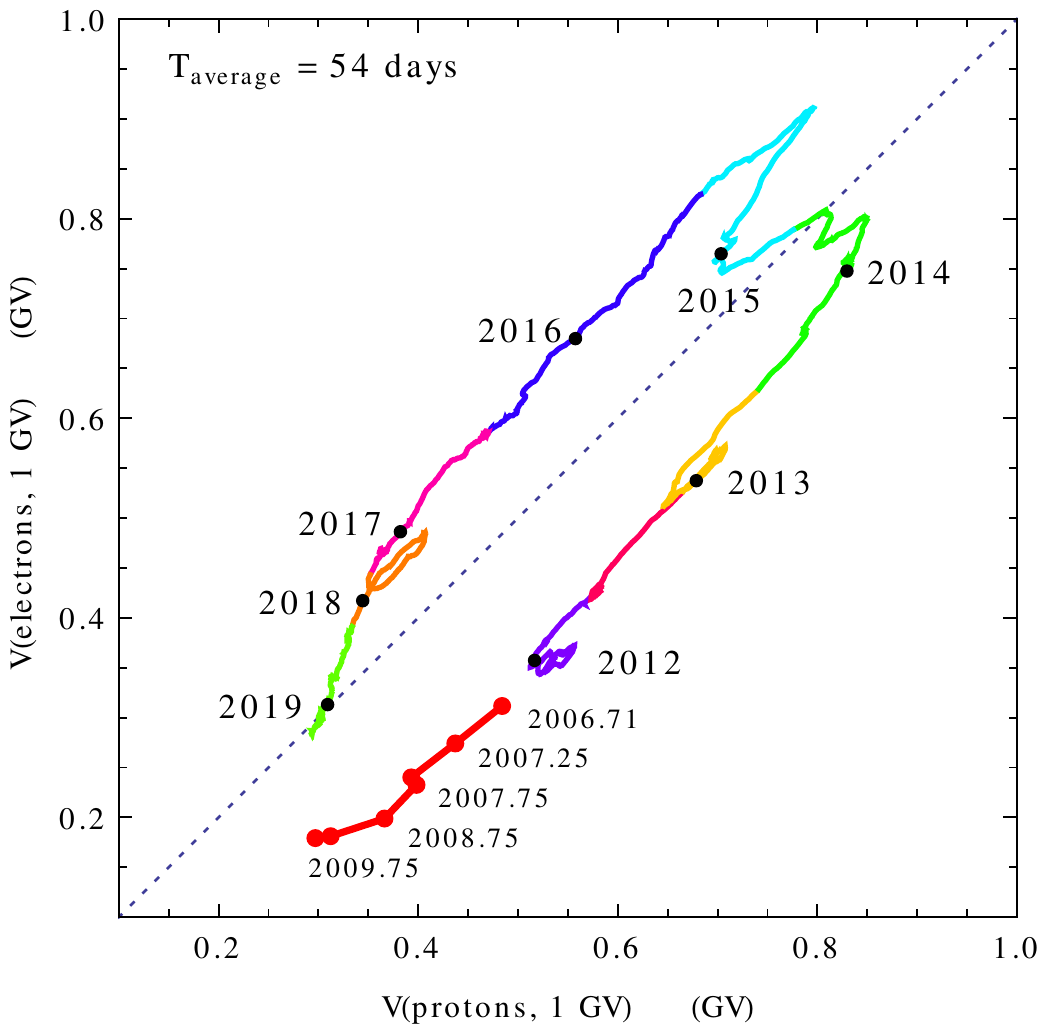}
\end{center}
\caption {\footnotesize
 Top panel:
 Trajectory of the point $\{V_p (t), V_{e^-} (t)\}$ that represent
 the potentials at rigidity 1~GV obtained fitting the $p$ and $e^-$ daily spectra
 measured by AMS02.
 The data of PAMELA cover the time interval from July 2006 to  December 2009,
 that is the final part of solar cycle 23. During this time
 the PAMELA collaboration has released seven electron spectra (large points in the plot).
 After a time gap, the AMS02 data start in May 2011 and end in October 2019.
 In the bottom panel the potentials for the AMS02 data
 are moving averages  with averaging time of 54 days and one day step
 of the potentials shown in the top panel.
 \label{fig:p-e_vcorr}}
\end{figure}

An instructive way to compare the proton and
electron potentials is shown in Fig.~\ref{fig:p-e_vcorr}, where the top
panel shows the trajectory
of the point $\{V_p(t), V_{e^-} (t)\}$ that represents the potentials at
rigidity 1~GV obtained fitting electron and proton spectra measured
at the same time $t$ by PAMELA or AMS02.
For the PAMELA data the plot shows the potentials obtained fitting the
seven electron spectra in
\cite{Adriani:2015kxa,Adriani:2016uhu}, together with an interpolation
of the potentials obtained fitting the proton spectra
\cite{Adriani:2013as,Martucci:2018pau}.
The PAMELA measurements are in the time
interval from July 2006 to October 2009 and cover the 
last part of solar cycle 23 when solar activity goes toward its minimum,
with polarity $A < 0$.
For the AMS02 data we show the results of fits to all days
where both $p$ and $e^-$ spectra have been measured, with the broken line
connecting all measurements (in order of increasing time).
The AMS02 data start in May 2011, and covers most of solar cycle 24,
including the phase of solar maximum where one observes
the reversal of the solar magnetic field polarity.

Inspecting Fig.~\ref{fig:p-e_vcorr} one can observe some striking features, 
with the trajectory of the potential pair $\{V_p(t), V_{e^-} (t)\}$ 
that draws a loop.
From the beginning of the AMS02 observations until the solar maximum
around the middle of 2014 (a period where $A < 0$),
both potentials $V_p(t)$ and $V_{e^-}(t)$,
averaging over fluctuations,
grow gradually at approximately the same rate but with  $V_{e^-} (t) < V_p (t)$.
The time interval 2014--2016 corresponds to an extended
solar maximum phase and shows an
evident double peak structure separated by a gap, a structure
that is also observed in other solar cycles.
During this phase of the cycle the proton
potential reaches its maximum before the potential for electrons.
In the subsequent phase of the cycle
(with positive polarity $A > 0$) both potentials
decrease, again at approximately the same rate, but the inequality
for the potential is reversed ($V_{e^-} (t) > V_p (t)$).

In the top panel of  Fig.~\ref{fig:p-e_vcorr} the complicated form of the line
that connects the potentials for the AMS02 daily spectra encodes
valuable information, but performing moving averages of the
two potentials allows to obtain the much simpler trajectory, 
shown in the bottom panel,  where the ``global loop''
of the trajectory is more clearly visible.

The data strongly suggests that the
point $\{V_p(t), V_{e^-} (t)\}$ travels along the loop
in a clockwise sense for cycles (like solar cycle 23) where
the magnetic field polarity at the start of the cycle
(that is at solar minimum) is positive, and in an anti--clockwise
sense for cycles (like solar cycle 24) where the situation is opposite.

In fact, in Fig.~\ref{fig:p-e_vcorr}, one can observe
that in the time interval where only the PAMELA data are available
both potentials decrease gradually (with $V_{e^-} (t) < V_p(t)$),
and the pair $\{V_{p} (t), V_{e^-} (t) \}$
completes (around the end of 2006) a clockwise loop at the solar minimum that separates
solar cycles 23 and 24. After a gap in the observations of approximately
1.6 years, the AMS02 become available during the growing phase
of solar cycle 24, and one observes a reversal of the trajectory
with both potentials growing (with $V_{e^-} (t) < V_p(t)$ as before) and therefore moving in an
anti--clockwise sense along the loop.

If this scenario is correct, during the current solar cycle (number 25)
that started around December 2019, one should observe the point that
represents the potential pair to move in clockwise sense along a loop
that during the initial phase of increasing solar activity has $V_{e^-} > V_p$.

\subsection{Solar activity transients}
In the bottom panel of Fig.~\ref{fig:p-e_vcorr} are also evident 
some loop--like  structures of shorter time scale,
that are in coincidence with  similar structures observed for the flux--flux correlations
of a single particle  (as  discussed in section~\ref{sec:flux_corr}
and illustrated in Fig.~\ref{fig:corr_particles}).

These effects can be also observed studying correlations
between the values of the potential at different rigidities.
This is illustrated in Fig.~\ref{fig:v-running} that shows the trajectory of the point
$\{V_1 (t), \Delta V(t)\}$ where $V_1 = V_{[1~{\rm GV}]} $ and
$\Delta V = (V_{[10~{\rm GV}]} - V_{[1 ~{\rm GV}]})$ that describes the time evolution
of the potentials obtained fitting the daily spectra 
measured by AMS02 for protons, helium nuclei and electrons.
In the three panels at the top the broken line
connects the results of the fits to the daily spectra for the three particle types
(the errors are not shown to avoid clutter).
One can note that for a fixed value of $V_1$, the value of $\Delta V$ is not unique but it
has a finite range.
The three panels in the middle row of 
the figure show moving averages of the potential
after integration over time intervals of 81 days (three Bartels rotations).
The simplification obtained performing
the moving average allows to make evident some interesting ``hysteresis structures''.
These structures are of course the same ones visible in the flux--flux correlations
of Fig.~\ref{fig:corr_particles}, it is however interesting to note
that the modulation potential describes the state of the heliosphere,
and is independent from the shape of the spectra of the particles
in interstellar space.

Inspecting Fig.~\ref{fig:v-running} one can see that the hysteresis effects for
protons and helium are approximately equal, while the effects for electrons,
while strongly correlated, are significantly different.
This can be understood noting that the same solar activity events, such as large
CME's, can perturb both of the (different) regions
of the heliosphere where protons and electrons are propagating,
resulting in effects on the $p$ and $e^-$ spectra that are correlated but not identical.

The three panels in the bottom row of Fig.~\ref{fig:v-running}
show the trajectories of the potentials for moving averages with
a long integration time of 378 days (14 Bartels rotations).
For all three particles ($p$, He and $e^-$) one can see 
some significant differences (with the same qualitative structure)
for the average potentials during phases of the solar cycle before and
after solar maximum.
This effect is the same that was observed by AMS02
in \cite{ams_daily_helium} studying the helium/proton ratio.
It is difficult to say at the moment what is the origin of the
effect, and if it is associated to the ensemble of the
solar transient events in the solar cycle under study,
or is related to the general properties of the 22--year solar cycle.

As already discussed, performing moving averages
(of fluxes as in Fig.~\ref{fig:corr_particles}, or of potentials
as in Fig.~\ref{fig:v-running}) 
allows the visualization of interesting structures in CR modulation,
but also erases valuable information encoded in the
evolution of modulations for time scales shorter than the averaging time. 

To illustrate this point in Fig.~\ref{fig:vp-labels}
we show again the detailed (day to day) trajectory of the potential parameters
$\{V_1(t), \Delta V(t)\}$ for protons, 
indicating few (seven) days that correspond to major solar events.
These events have also resulted in Forbush decreases observed by neutron monitors.
To each events corresponds an large increase in the modulation potentiala,
and remarkably the increase of the potential at the higher rigidity (10 GV) is stronger
than at the lower one (1~GV).
These effects (as discussed in section~\ref{sec:flux_corr}) have been revealed in the past
\cite{rajan_loops,Munini:2018cgc,Lagoida:2021udw,DAMPE:2021qet},
but a detailed explanation is still under construction.

\begin{figure}
\begin{center}
\includegraphics[width=9.0cm]{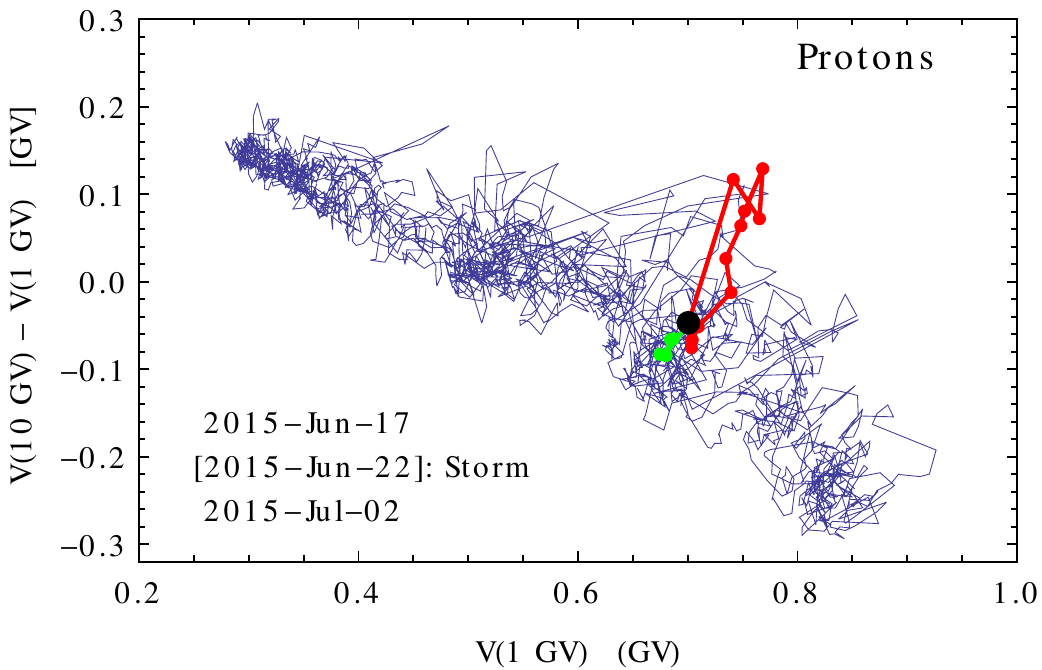}

\vspace{0.8 cm}
\includegraphics[width=9.0cm]{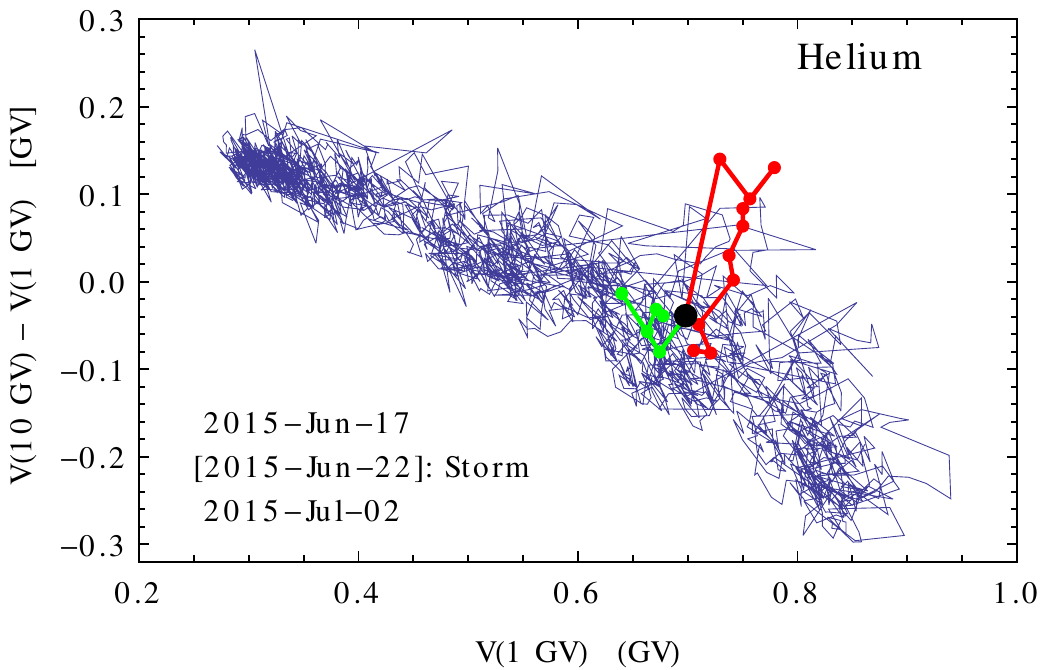}

\vspace{0.8 cm}
\includegraphics[width=9.0cm]{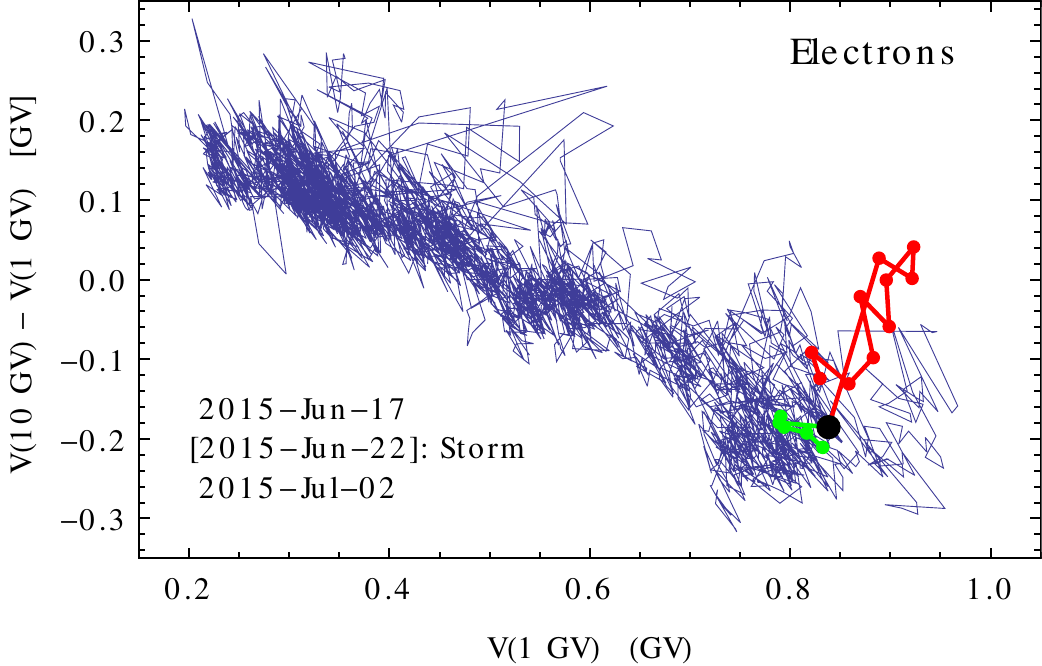}
\end{center}
\caption {\footnotesize
 Trajectory of the point $\{ V_1(t), V_2(t) \}$ during the solar event
 around 2015-Jun-22. The gray line connect fits to all  the daily spectra.
 The large black point corresponds to
 the day 2015--Jun--22 when a large solar storm was detected at ground level.
 The green points show the potentials for the 5 days before the storm
 (starting 2015--June--17). The red points show the potentials for
 10 days after the solar storm (ending 2015--July--02).
\label{fig:event1}}
\end{figure}

The effects of large solar activity events on the CR spectra can evolve very
rapidly on a time scale of hours, and following the details of this evolution,
can be of great help to develop an understanding of these phenomena.
the AMS02 daily measurements are therefore of great interest.
As an example,  in Fig.\ref{fig:event1} we show the the  trajectory
of the modulation potentials (for $p$, He and $e^-$) obtained fitting
the AMS02 daily spectra obtained during few days around
one of the largest solar events during solar cycle 24.
This event was observed around the
summer solstice of 2015 \cite{event-summer-solstice-2015}.
From 18--23 June, one of the largest sunspot active regions in the Sun (AR 12371),
at the time directly facing Earth, produced several flares,
giving origin of four CME impacting Earth in the period 21-25 June.
The third and largest impact (June 22nd) 
generated a G4-severe geomagnetic storm with spectacular auroras even
at low latitudes, followed by a Forbush decrease observed by ground-level detectors.
Fig.~\ref{fig:event1} puts in evidence the trajectories of the modulation
potential for $p$, He and $e^-$ (represented by the pair $\{V_1(t), \Delta V(t)\}$)
taken during a time interval of 16 days around the date of the
solar storm (starting 5 days before, and ending 10 days after).
A detailed description of this event is not possible here, but one can note
that it generated distortions of the spectra for all three CR particles
of very similar structure.
The spectral distortions generated by the event
developed rapidly, with a time scale of one day or less;
following this, the spectra returned to their pre-solar-event values
with a longer time scale of several days.
As noted before, the distortions (measured by the variation of the modulation potential)
were larger at the higher rigidity of 10~GV, and weaker at $\rig \simeq 1$~GV, and this 
appears to be the case in most if not all cases.
Time structures qualitatively similar to what we have described
can be observed for other large solar activity events. 

\section{Summary and Conclusions}
Most of the already rich literature that discusses the PAMELA and AMS02
data  is based on the study of the time dependence of the CR fluxes in
different intervals of rigidity (or energy).
An alternative possibility,  it to  extract from the data some
time dependent parameters that describe the CR spectral shapes.
In this work we have   used this  second approach, and  demonstrated
that it is possible to accurately and economically describe the  CR spectra
of each particle type in terms of a time dependent  modulation potential $V (\rig, t)$.
A small (but not negligible)  rigidity dependence of the potential is required
to fit the high precision data that are now available.

The main goal of this work has been to investigate
the  origin of the phenomena  observed by the AMS02 collaboration
and  called   ``hysteresis effects''. Two of such effects have been shown by
combining measurements of the fluxes of helium and protons
\cite{ams_daily_helium} and
of electrons and protons \cite{ams_daily_electrons}.
We suggest that two distinct mechanisms are acting to generate the effects.

A first mechanism is at the origin of the largest effect,
that is observed for the $e^-/p$ combination with a long ($\gtrsim 1$~yr)
time scale. This effect can be described as an ``hysteresis loop''
for the potentials  $V_p (t)$ and $V_{e^-} (t)$ (at any  fixed rigidity $\rig$)
for $p$ and $e^-$ spectra, with the same period of the 11--year solar cycle
(note that these loops can also be  observed
as the hysteresis of the fluxes $\{\flur_{p} (\rig,t), \flur_{e^-} (\rig, t)\}$).

In fact, we suggest that this effect generates a ``double loop''
with the potentials moving along trajectories of similar form
but in opposite directions in alternate solar cycles.
Fitting the AMS02 data one observes that
during the first part of solar cycle 24, before maximum,
the two potentials $V_{p} (t)$ and  $V_{e^-} (t)$,
after averaging over fluctuations generated by solar transient events,
increase gradually with solar activity with $V_{p} (t)> V_{e^-} (t)$.
After solar maximum the two potentials decrease gradually,
but the inequality is reversed: $V_{p} (t) < V_{e^-} (t)$.
This  results in a trajectory of the point $\{V_{p} (t), V_{e^-} (t)\}$
that follows, in an anti--clockwise sense,
a loop--like trajectory.

There are indications from the PAMELA data that a similar trajectory,
but moving in the opposite direction,  was followed by the $p$ and $e^-$ potentials
during the previous solar cycle that finished in December 2009.
It is now natural to predict that the pair of $p$ and $e^-$ potentials will move
along loops of similar form, in opposite senses during even and odd solar cycles.

This prediction is based on some simple and well established results about the
propagation of charged particles in the heliosphere.
Because of the structure of the regular solar magnetic field
one has that when $q A > 0 $ (that is when the product
of the electric charge $q$ of the cosmic rays
and the polarity $A$ of the solar magnetic field is positive)
the CR particles arrive at the Earth mainly from the
heliospheric poles,
while in the opposite case ($q A <0$) the CR particles arrive mainly 
along the current sheet near the heliospheric equator.
The energy loss suffered by the particles during propagation
(and therefore the size of the modulations) at the same phase in a cycle
is larger for propagation close to the current sheet, and therefore
one has the inequality
\begin{equation}
V_{[q A > 0]}(t) < V_{[q A < 0]}(t) ~.
\end{equation}
The polarity $A$ is reversed at solar maximum (in the middle of one solar cycle),
and this, combined with the fact that the potentials are correlated with
the 11-year cycle of solar activity, generates the double loop structure.

A second mechanism is at the origin of two other effects
discussed in the AMS02 publications, namely: \\
(i) the ``sharp structures''
observed in the $e^-/p$ hysteresis that correspond to structures
observed in the time evolution of the fluxes for both particle types
\cite{ams_daily_electrons}.
Similar sharp structures have not been reported but  are also present
for He/$p$  hysteresis curves, and  become evident performing moving averages
with  integration times of 10--100 days.  \\ 
(ii) the hysteresis effects observed combining the proton and helium fluxes
\cite{ams_daily_helium}.

In this work we argue that both effects (i) and (ii)  have their
origin in the fact that CR spectra at the Earth
suffer modulations that cannot be described by one family of curves
controlled by a single time dependent parameter,
because the distortions generated by modulations can have different shapes at different times.
More explicitely, the value of the spectrum at one rigidity $\rig_1$ does not determine
uniquely the spectrum at a different rigidity $\rig_2$.

These variations in spectral shape can be observed 
studying the hysteresis of pairs of measurements such as
$\{ \flur (\rig_1,t), \flur (\rig_2, t)\}$  of the flux of a single particle type
for two distinct values of the rigidity,
or alternatively of the hysteresis for pairs of potentials
$\{ V (\rig_1,t), V (\rig_2, t)\}$.

These studies reveal that solar activity events, like large 
CME's, that perturb the heliosphere causing rapid variations
(or ``sharp structures'') in the time evolution of the CR fluxes at
any (sufficiently low) fixed value of the rigidity,
generate spectral distortions that are rigidity dependent,
with effects that are in general
more rapid and stronger at higher $\rig$.
Therefore an hysteresis curve $\{ \flur (\rig_1,t), \flur (\rig_2, t)\}$
or $\{ V (\rig_1,t), V (\rig_2, t)\}$
in the presence of one such transient will also exibit a ``sharp structure'', typically 
in the form of a clockwise (for $\rig_2 > \rig_1$) loop that extends
for the duration of the heliospheric perturbation associated to the
solar transient.

These effects  are also visible in hysteresis studies,
such as those performed by AMS02, that combine measurements of the fluxes
of different particles at the same rigidity.
This is the case when comparing protons and electrons, when 
(as discussed above) the particles suffer different modulations,
but it also true comparing protons and helium nuclei,
that suffer modulations that are approximately equal, 
because the modulations effects act as distortions
on LIS spectra that have different shapes.

On the other hand, if the study is performed for the modulation potentials
(that are independent from the shape of the LIS spectra)
the sharp loop--like structures associated with solar activity events
are absent for the hysteresis  of the  potentials of
$p$ and He, because the two particles suffer approximately equal  modulations,
while they continue to exist for the $p/e^-$ comparison,
because the two particles types have opposite electric charge and
propagate in different regions of the heliosphere, that are disturbed in different ways
by the solar events.

\begin{figure}
 \begin{center}
\includegraphics[width=5.0cm]{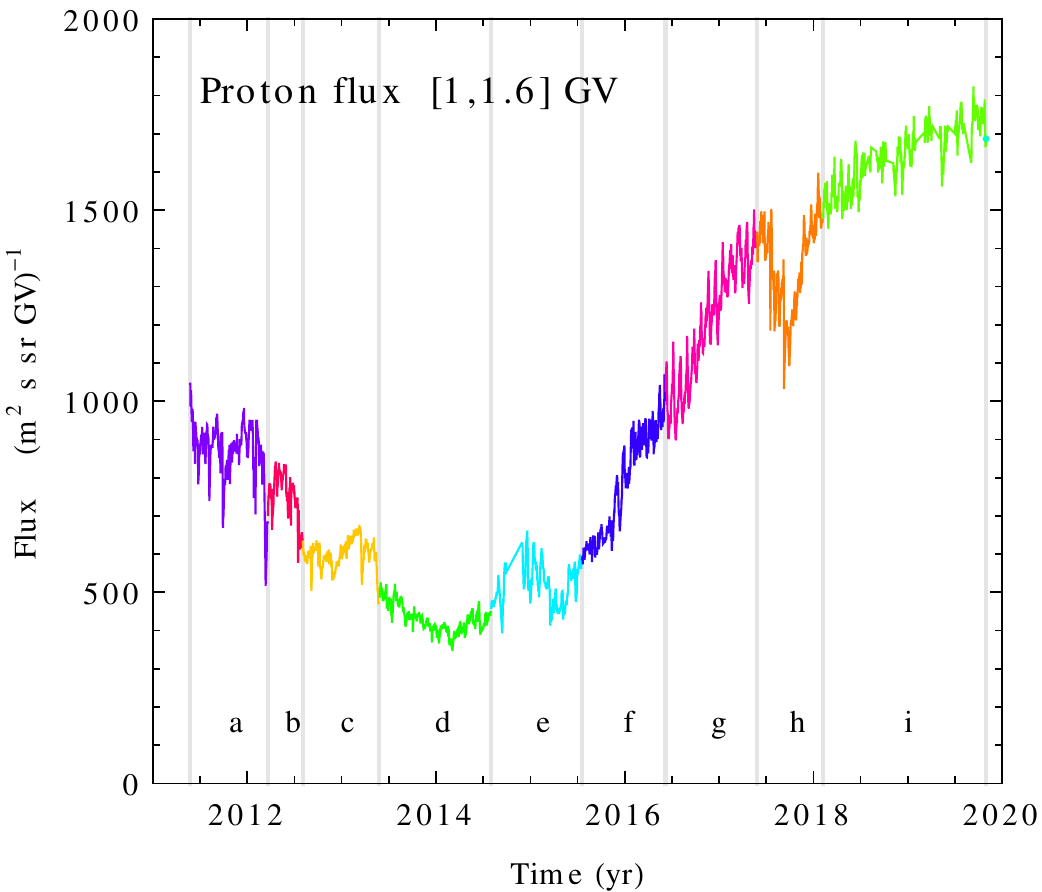}
~~
\includegraphics[width=5.0cm]{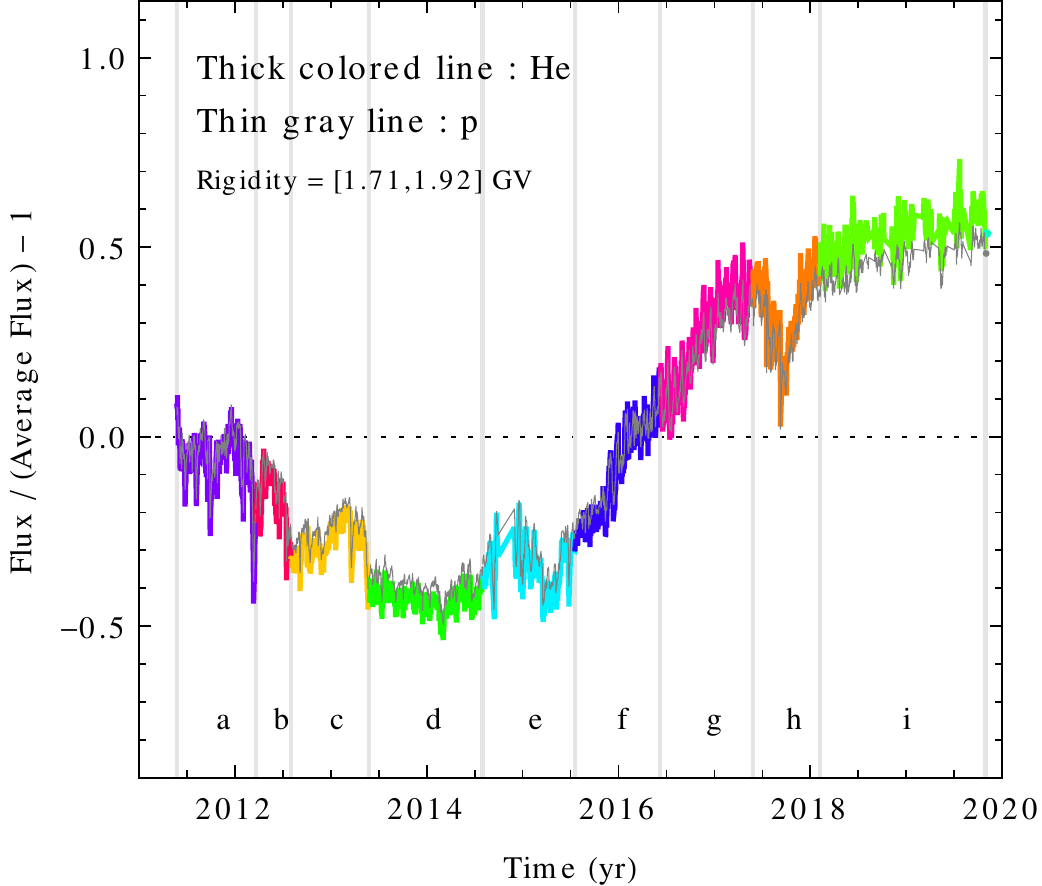}
~~
\includegraphics[width=5.0cm]{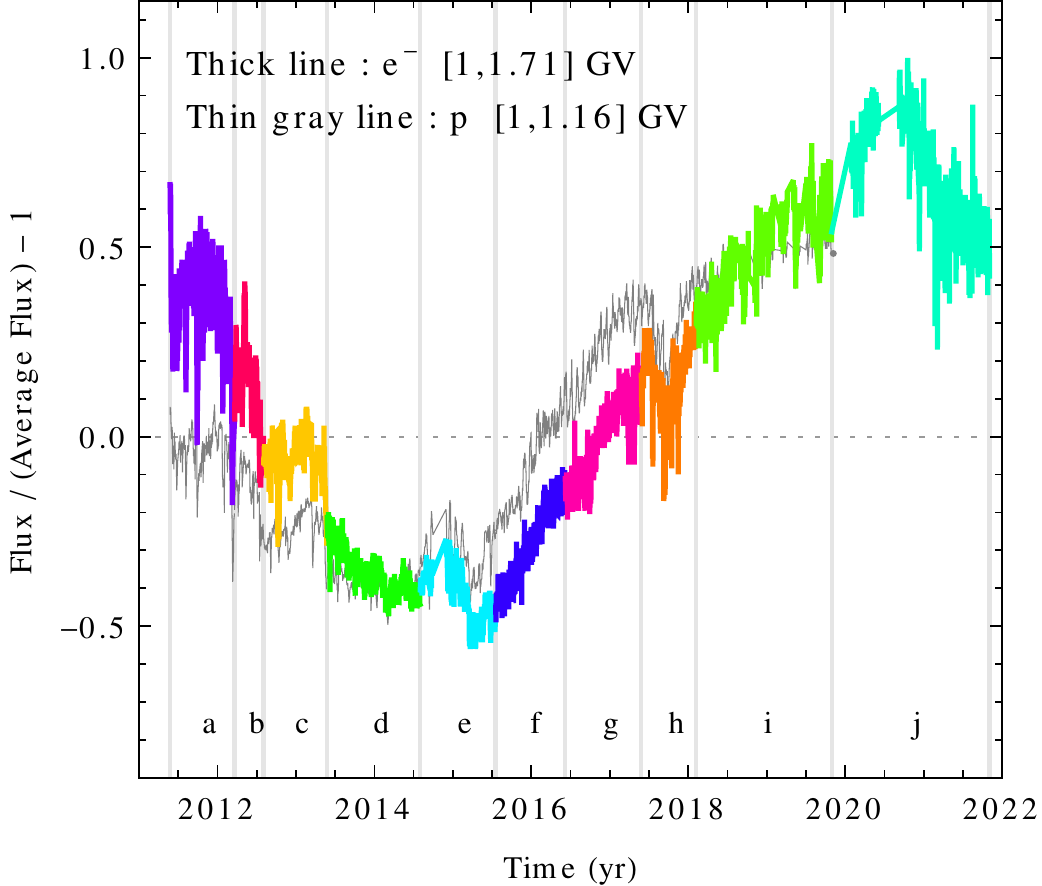}

\vspace{0.4 cm}
\includegraphics[width=5.0cm]{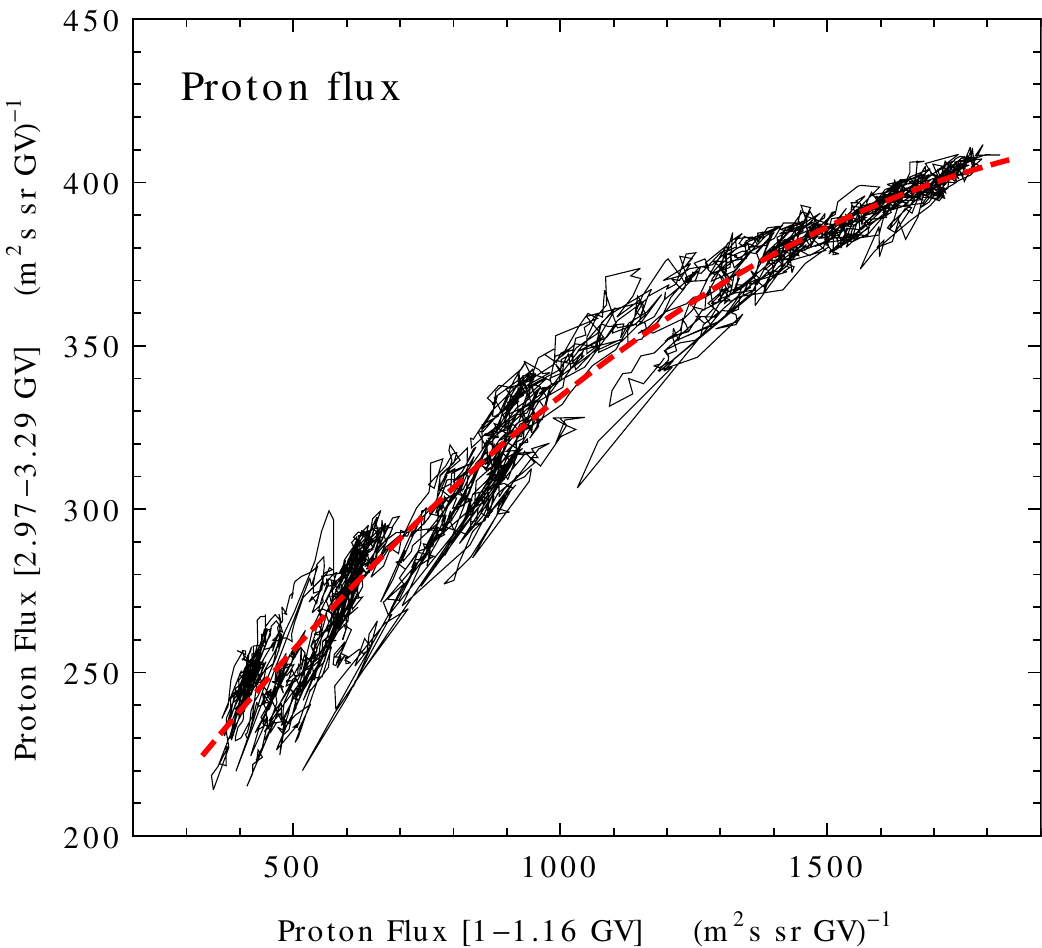}
~~
\includegraphics[width=5.0cm]{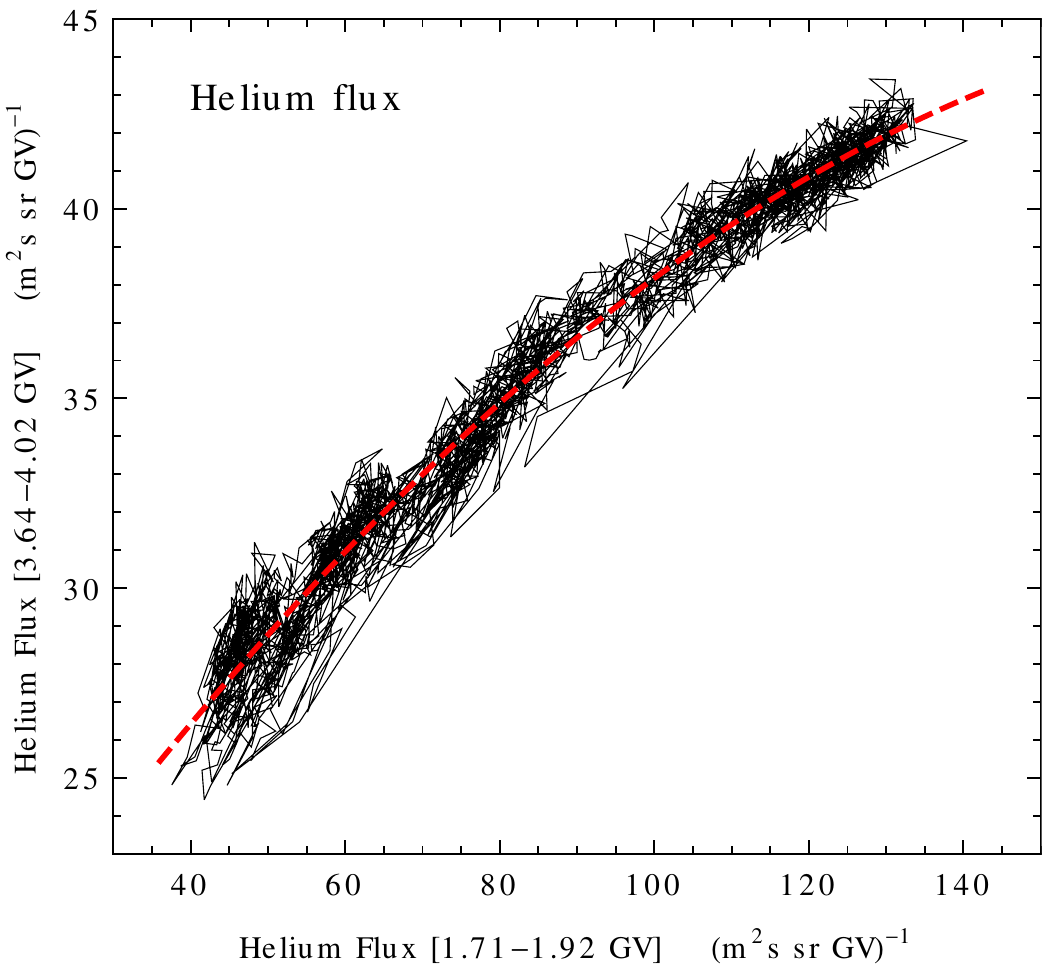}
~~
\includegraphics[width=5.0cm]{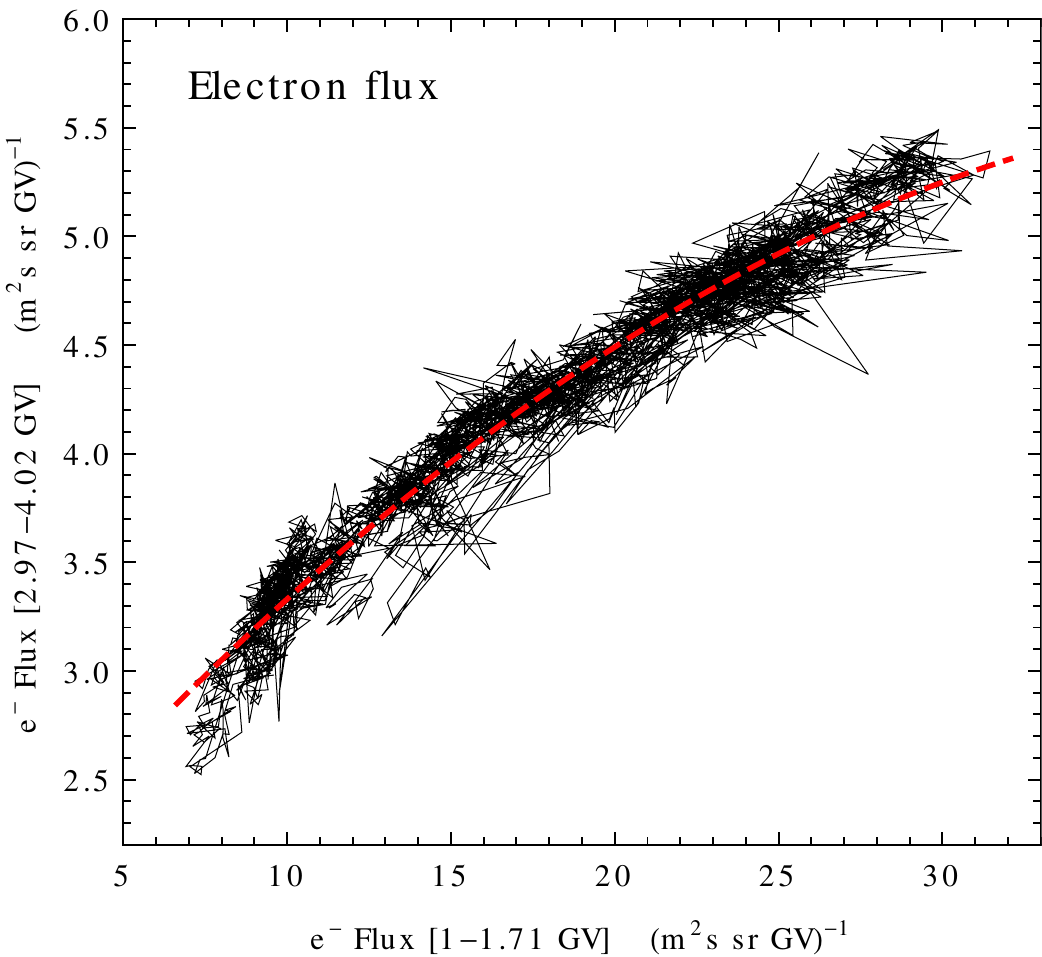}

\vspace{0.4 cm}
 \includegraphics[width=5.0cm]{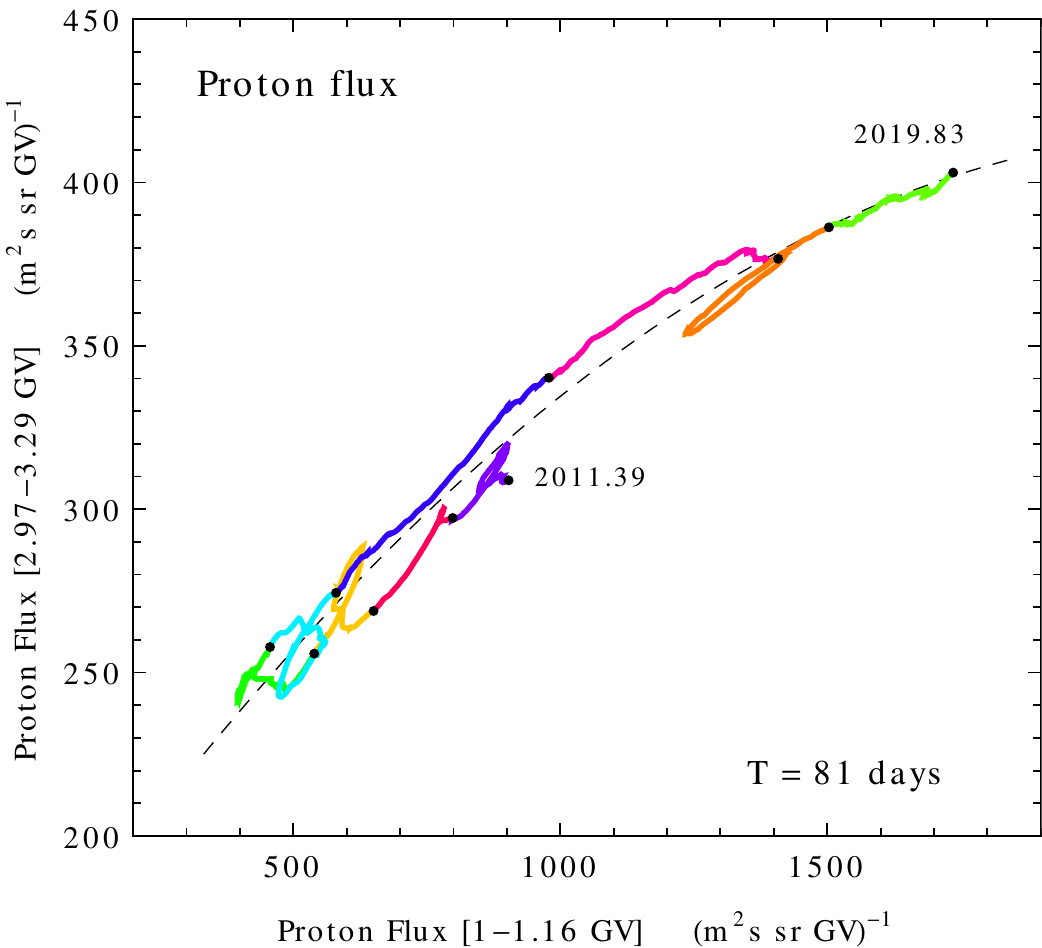}
~~
 \includegraphics[width=5.0cm]{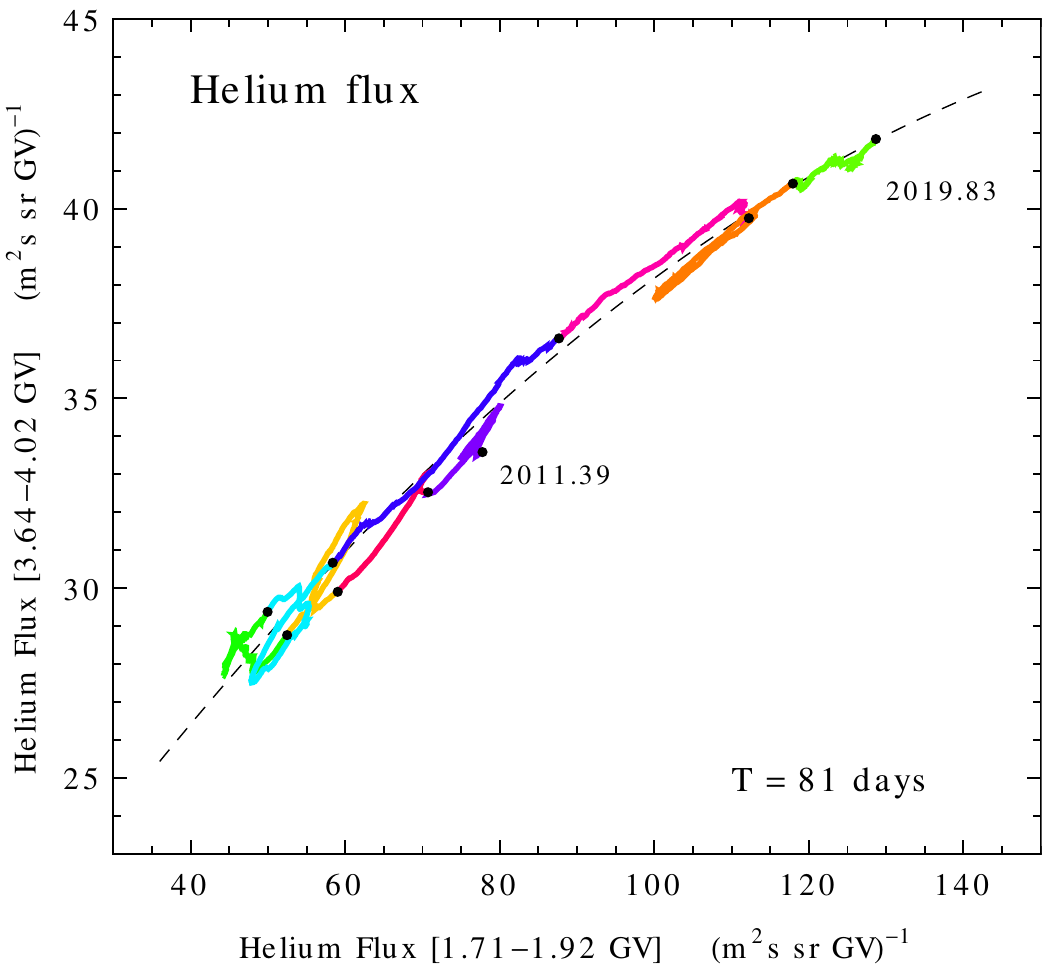}
~~
 \includegraphics[width=5.0cm]{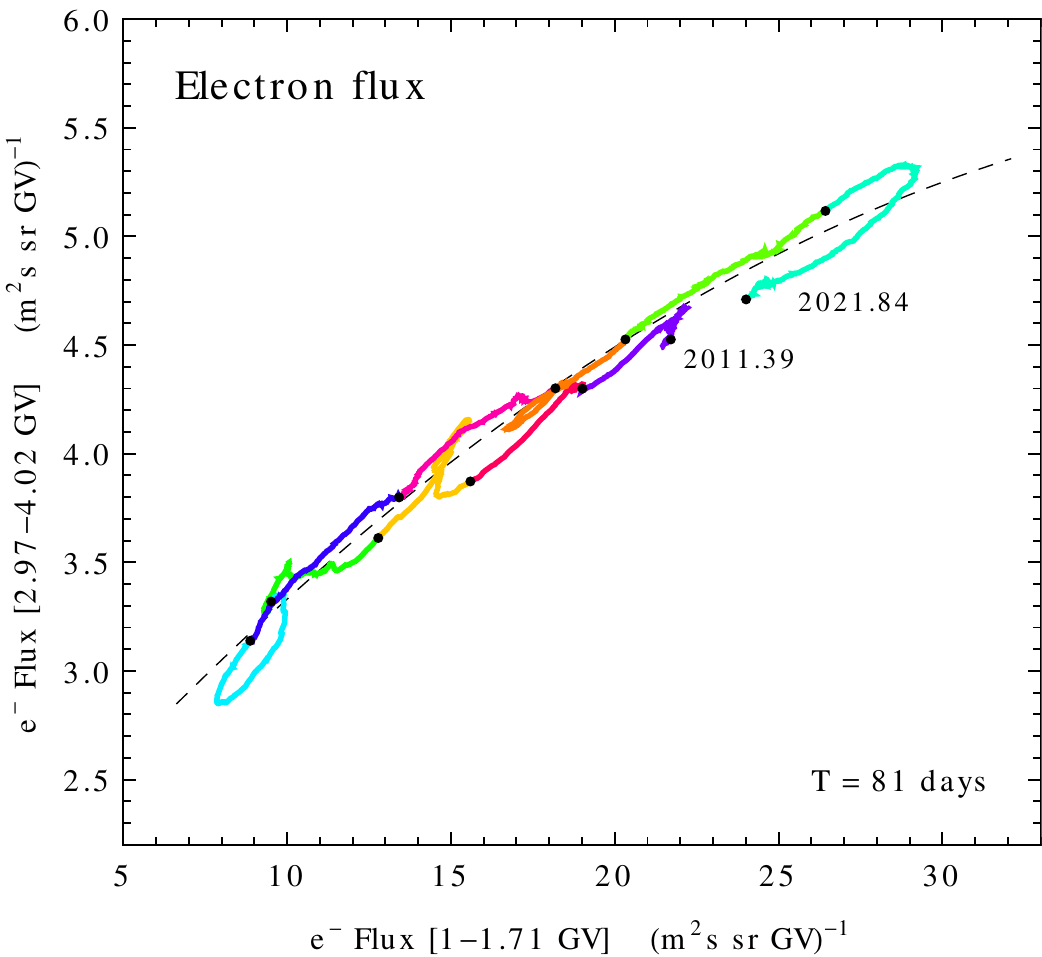}

\vspace{0.4 cm}
\includegraphics[width=5.0cm]{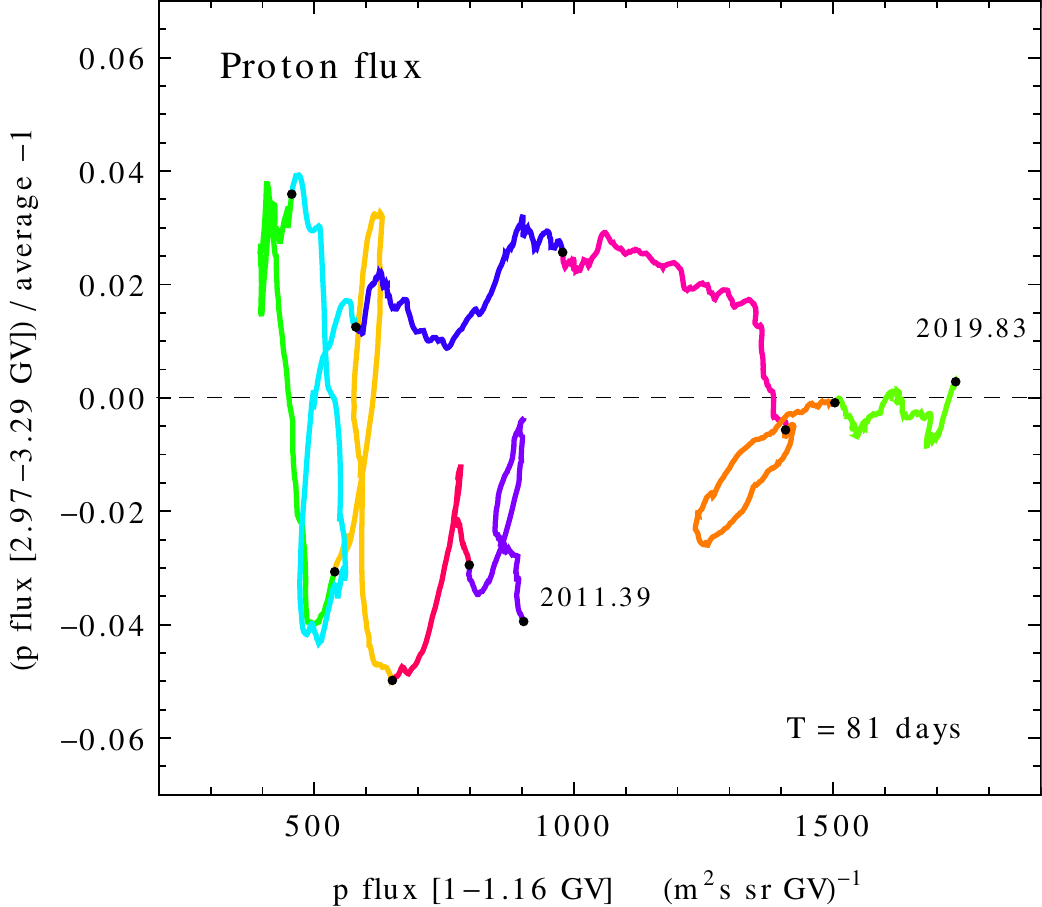}
~~
\includegraphics[width=5.0cm]{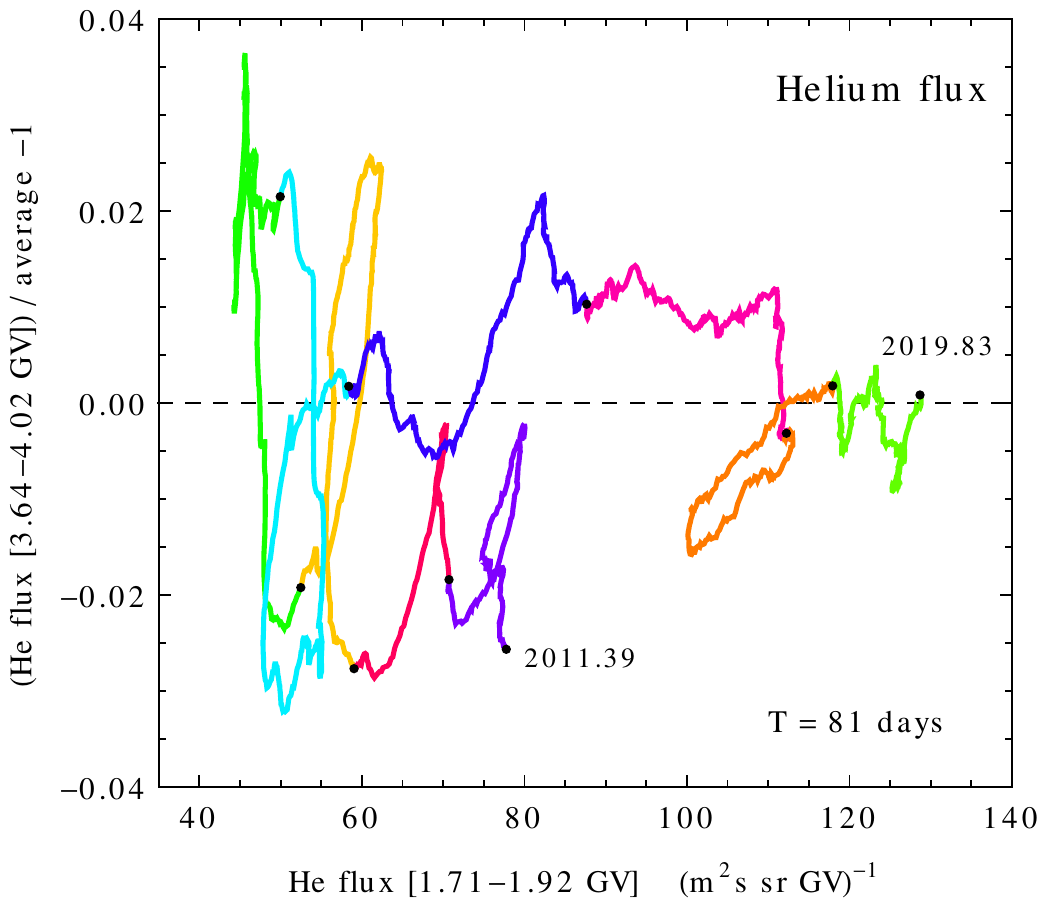}
~~
\includegraphics[width=5.0cm]{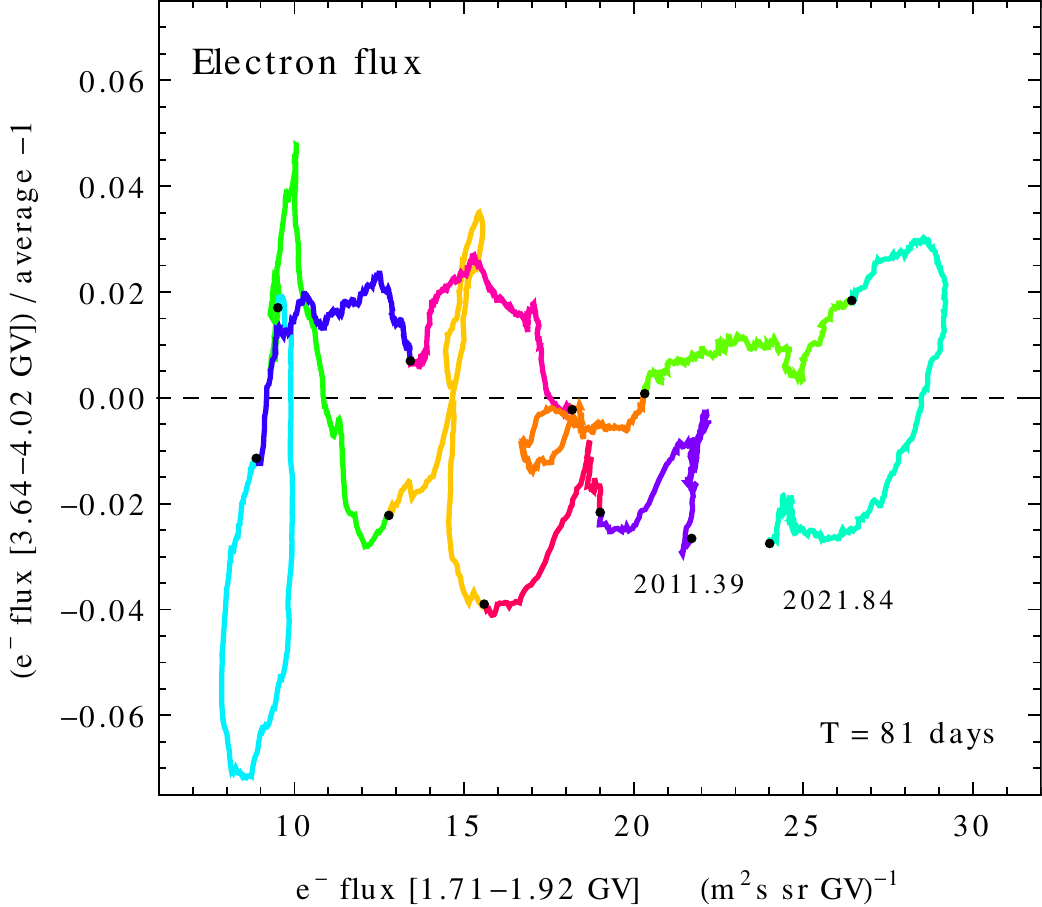}

\end{center}
 \caption {\footnotesize
The three panels in the top row show the time dependence
of the daily flux of protons 
(rigidity interval $\rig =[1,1.16]$~GV),
helium nuclei ($[1.71,1.92]$~GV) and
electrons ($[1,1.71]$~GV).
Error bars (in all panels) are not shown to avoid cluttering. 
The colors and vertical lines identify different
time intervals of interest.
The panels for helium and electrons also show (as a gray line)
the time dependence of the $p$ spectrum for comparison.
The panels in the three lower rows
show the trajectories (as a function of time)
of the point 
$\{\flur_{[\rig_1, \rig_2]} (t),
\flur_{[\rig_1^\prime, \rig_2^\prime]} (t) \}$
that represents the measurements
of the CR flux in the rigidity intervals
$[\rig_1, \rig_2]$ and
$[\rig_1^\prime, \rig_2^\prime]$.
The three columns are for protons, helium nuclei and electrons.
In the panels of the second row a broken line connects
measurements taken in different days.
The panels in the third row show running averages
taken integrating the fluxes 
during a time interval of 81 days (3 Bartels rotations)
with colors indentifying the same time intervals of the top row.
The three panels in the bottom row show the same 
moving average as those above with the $y$ axis giving
the deviation of the flux $\flur_{[\rig_1^\prime, \rig_2^\prime]} (t)$
from the average value (indicated by a dashed line in the previous plots).
\label{fig:corr_particles}}
\end{figure}

\begin{figure}
\begin{center}
\includegraphics[width=5cm]{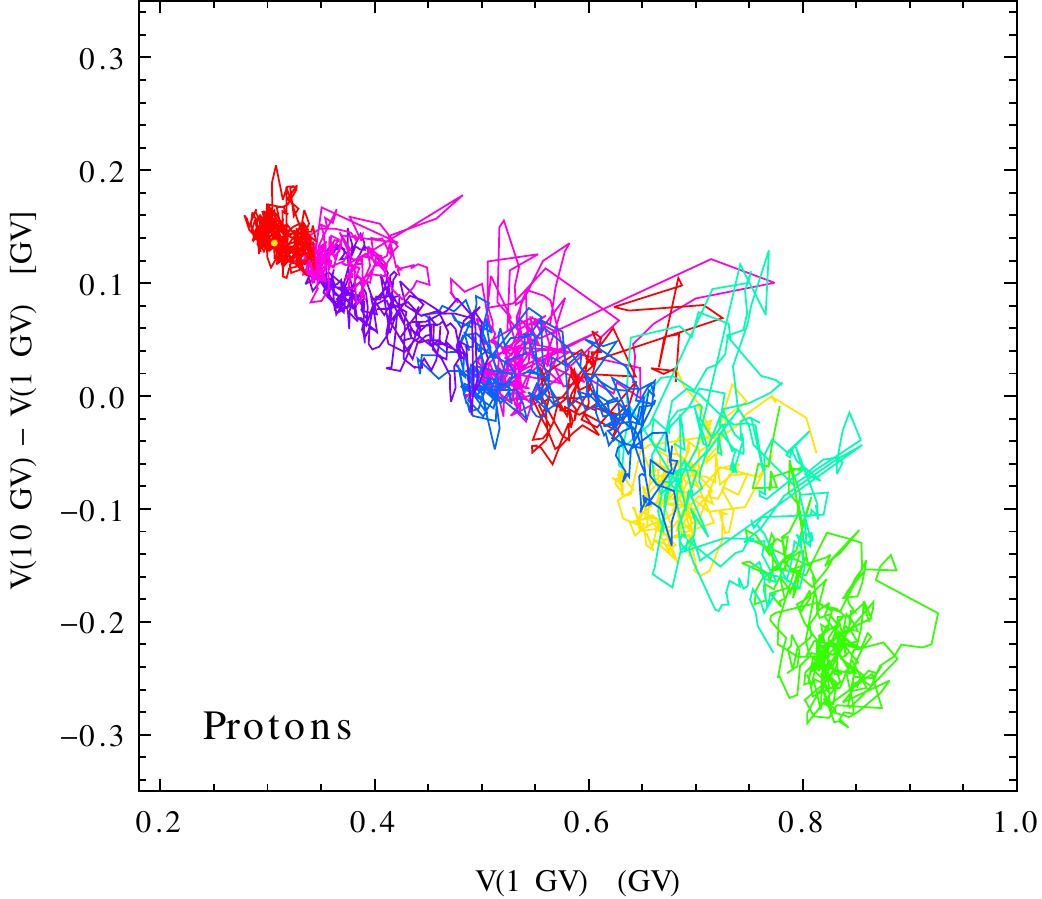}
~~
\includegraphics[width=5cm]{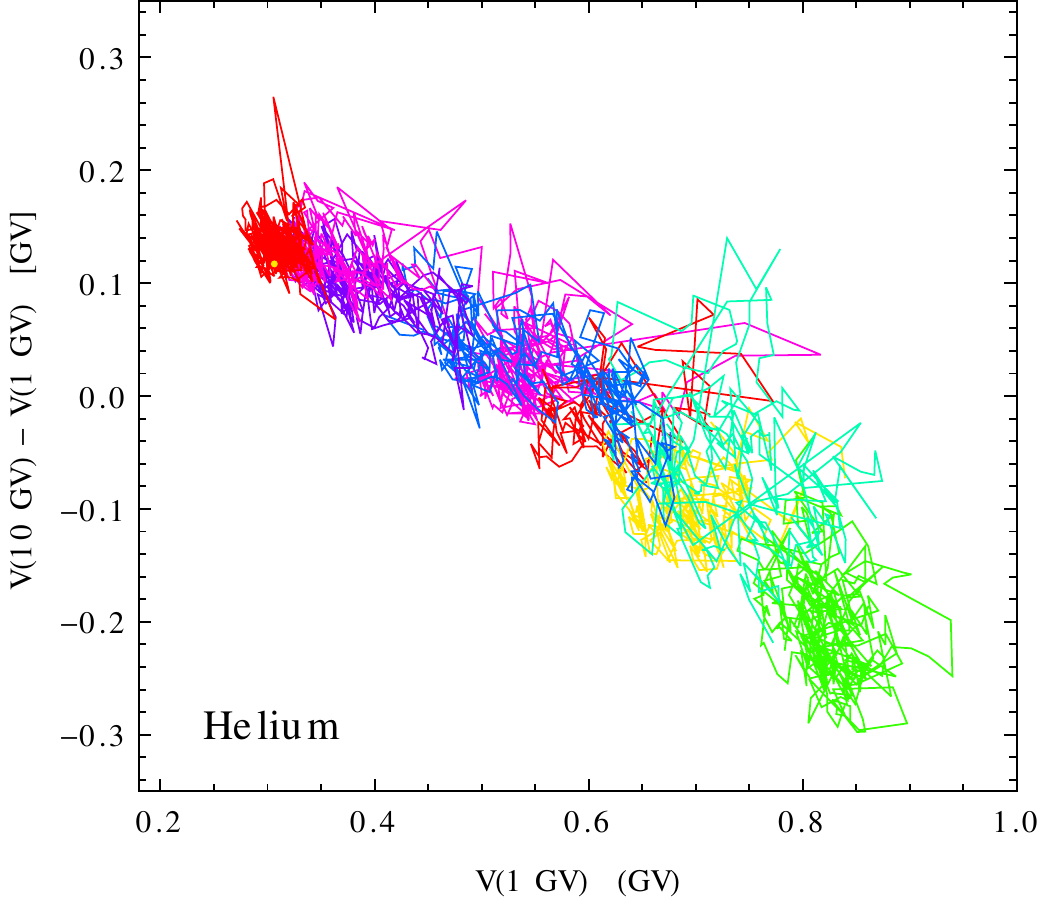}
~~
\includegraphics[width=5cm]{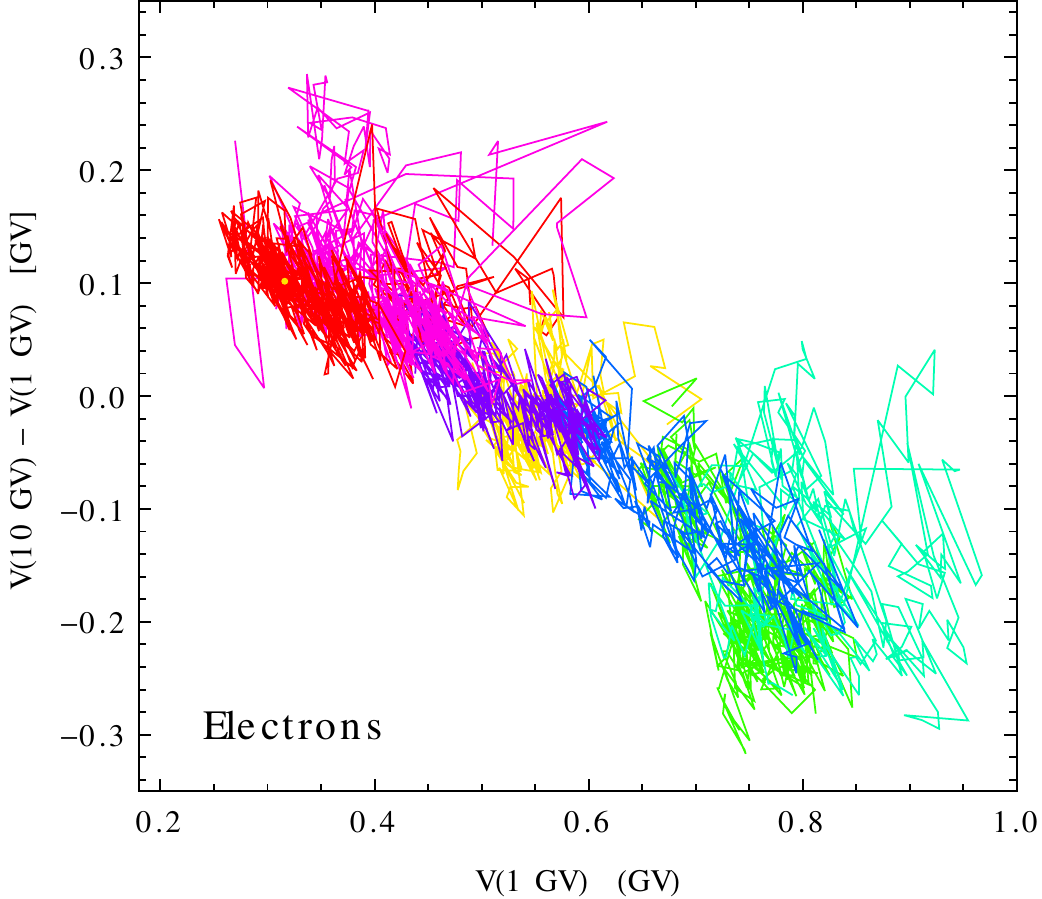}

\vspace{0.7 cm}
\includegraphics[width=5cm]{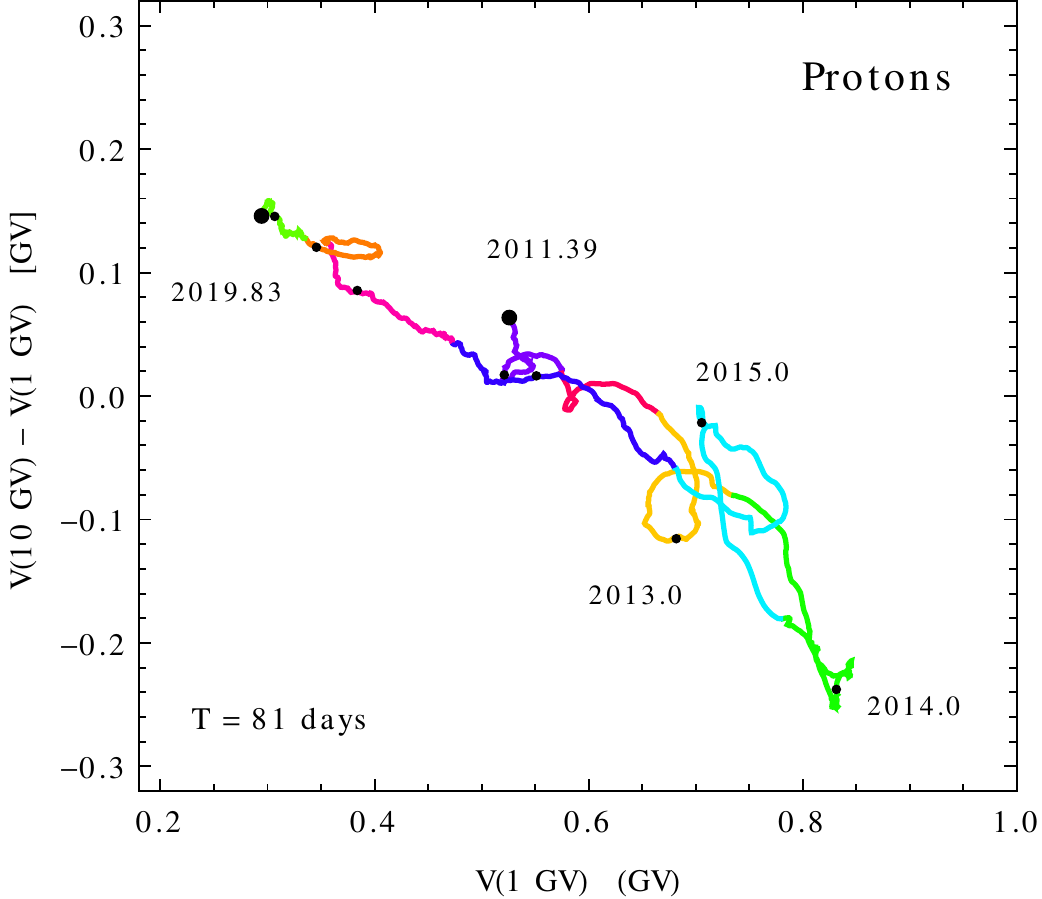}
~~
\includegraphics[width=5cm]{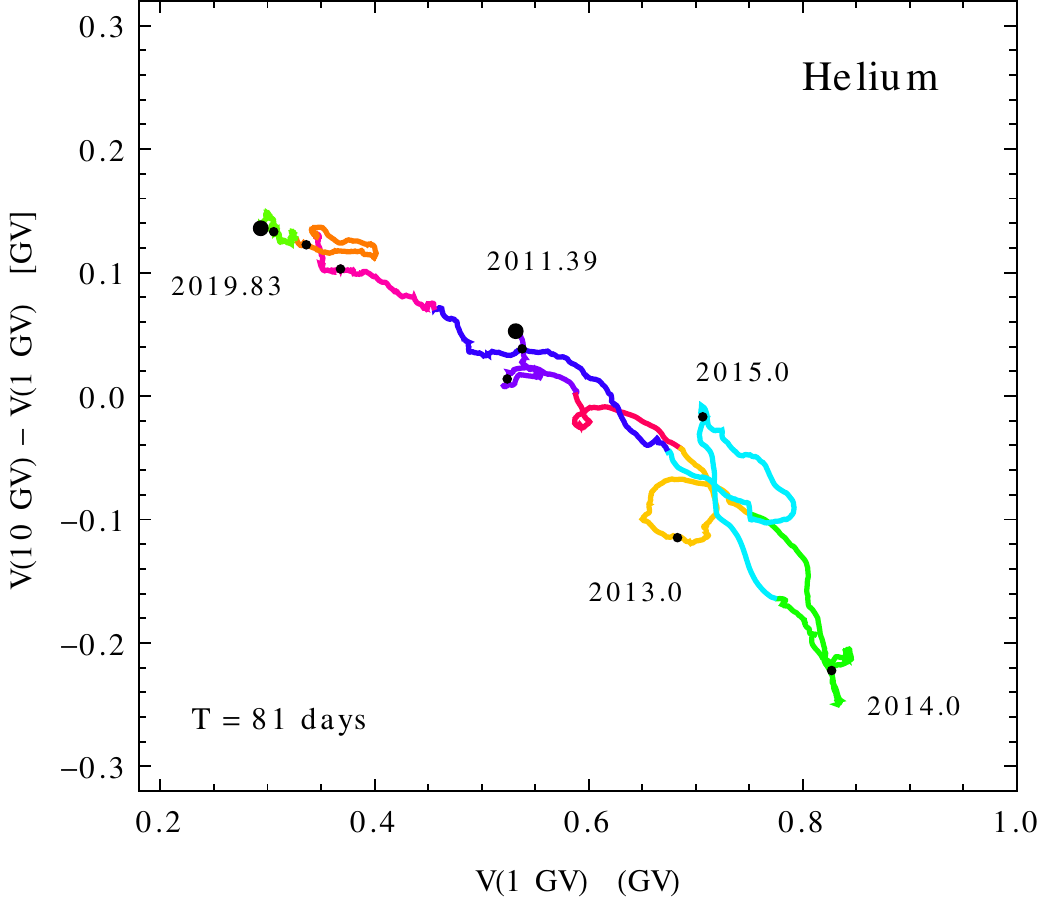}
~~
\includegraphics[width=5cm]{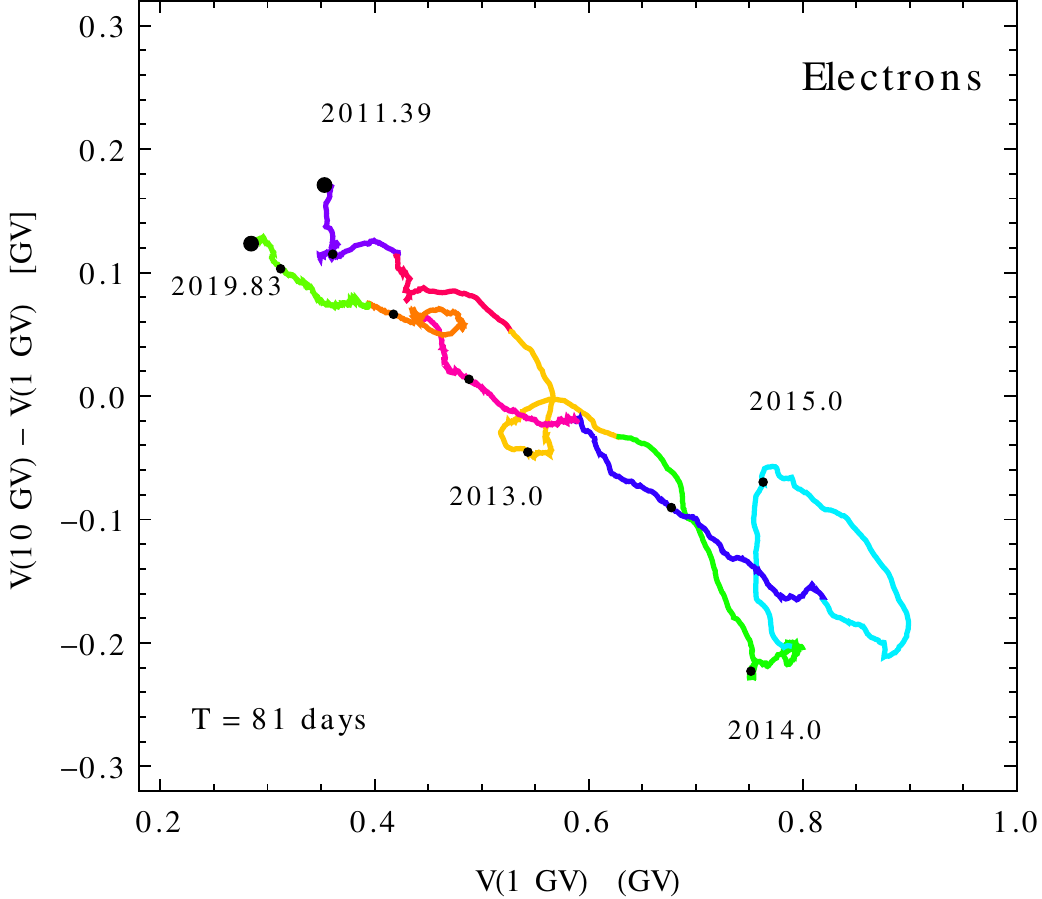}

\vspace{0.7 cm}
\includegraphics[width=5cm]{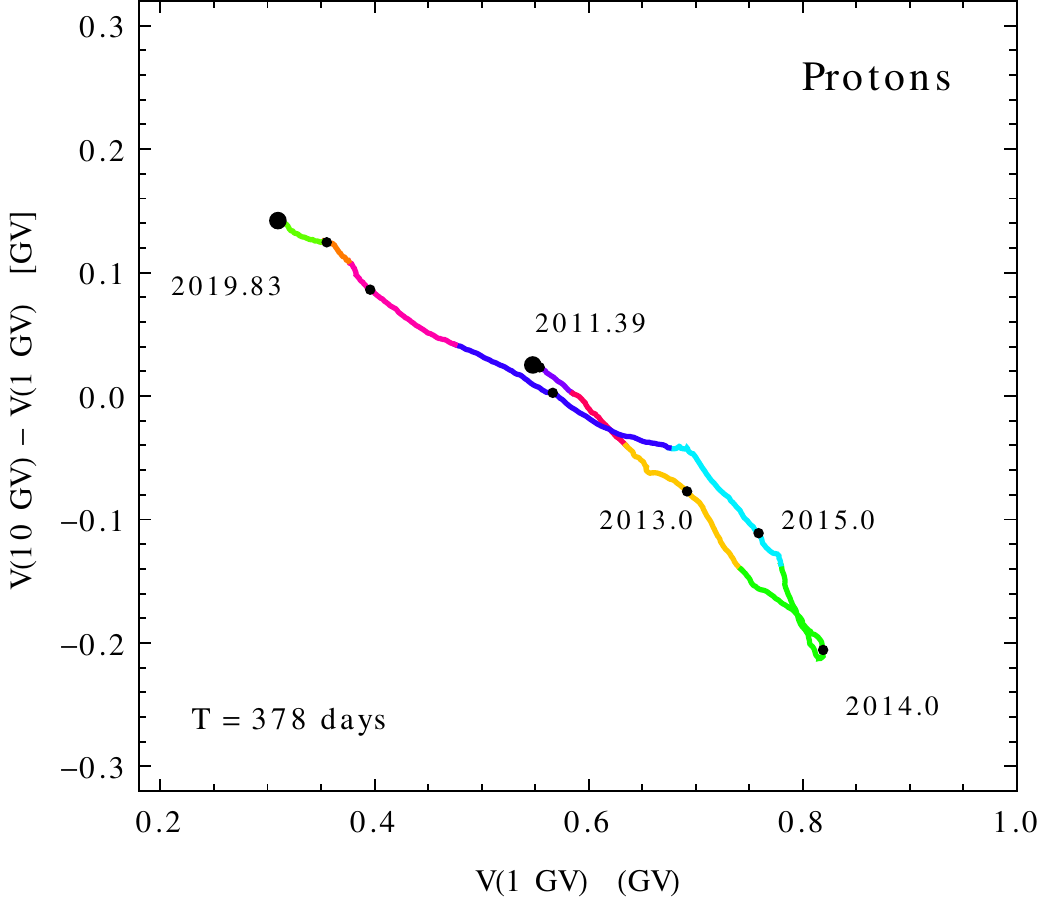}
~~
\includegraphics[width=5cm]{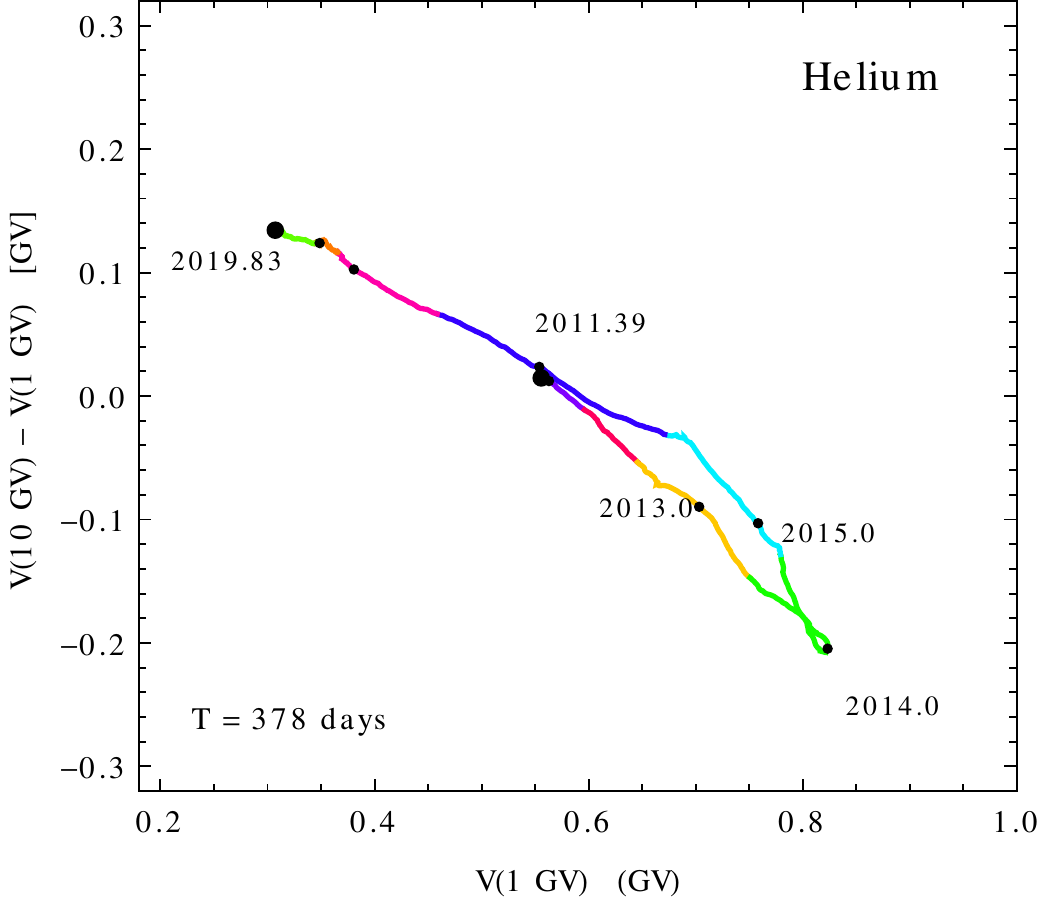}
~~
\includegraphics[width=5cm]{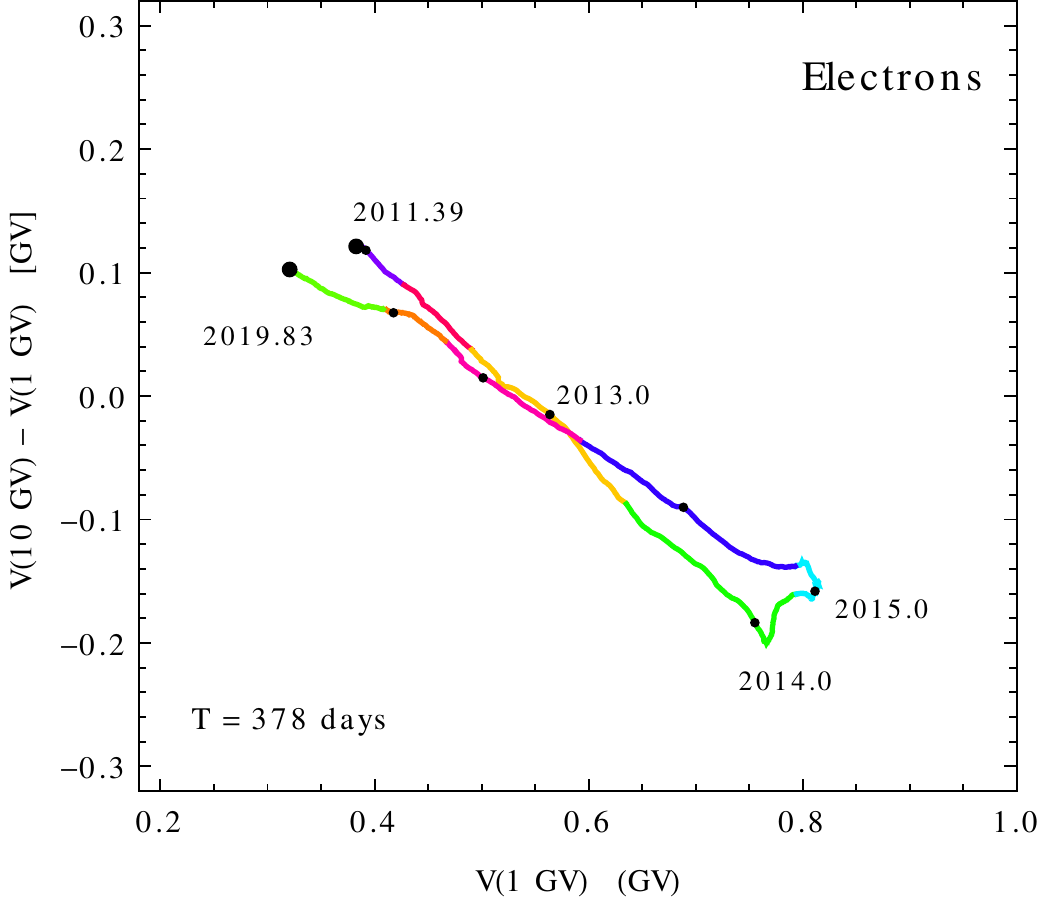}
\end{center}
\caption {\footnotesize
 The three panels at the top show the trajectory of the point
 $\{V_1 (t), \Delta V(t)\}$
 (with $V_1 = V_{[1~{\rm GV}]}$ and
 $\Delta V = V_{[10~{\rm GV}]} - V_{[1~{\rm GV}]}$)
 for protons, helium nuclei and electrons.
 The potentials are obtained fitting the AMS02 daily spectra.
 The three panels in the middle (bottom) show moving averages
 of the potentials calculated for an integration time of 81 days (378 days).
 The colors of the lines are the same used in Fig.~\ref{fig:corr_particles} and
 identify the same time intervals.
 \label{fig:v-running}}
\end{figure}

\begin{figure}
\begin{center}
\includegraphics[width=11cm]{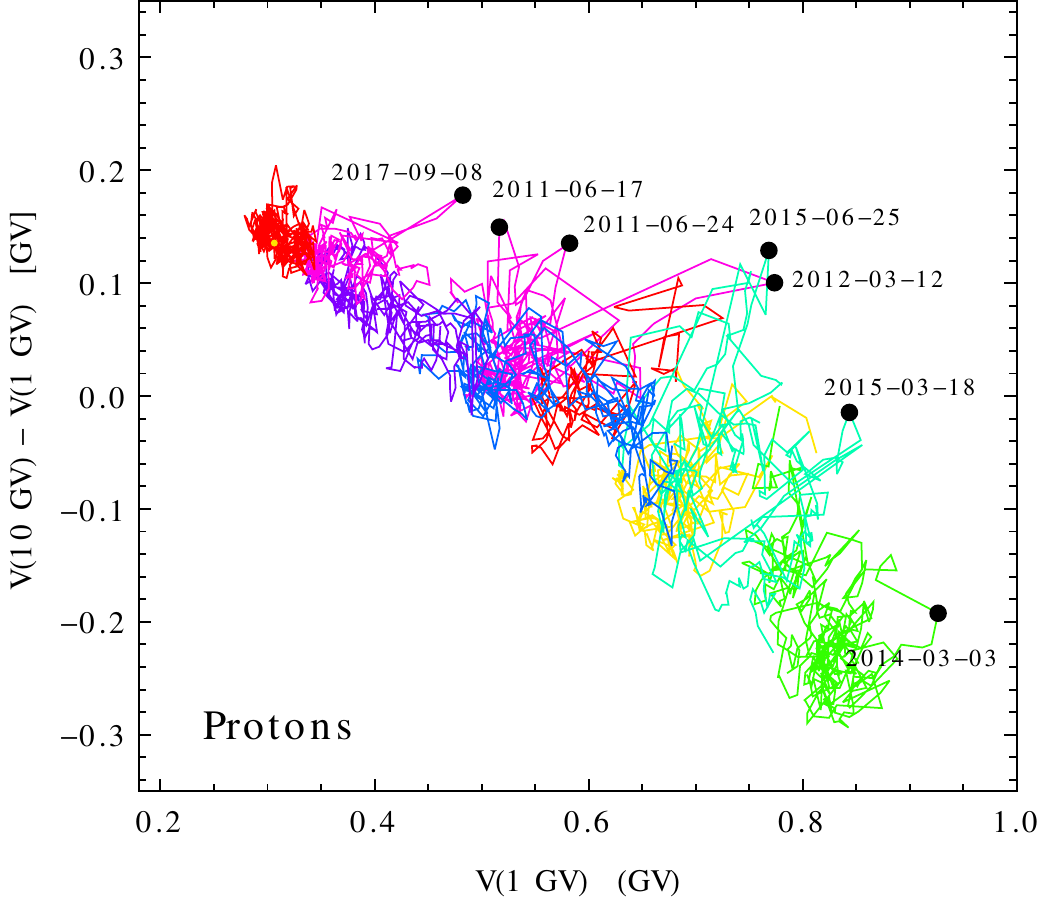}
\end{center}
\caption {\footnotesize
 Trajectory of the potentials for protons (as in top--left panel
 of Fig.~\ref{fig:v-running}. The dots indicate the
 dates where the shape of the spectrum is most distorted.
 All these dates can be associatewd  to large solar activity events and to
 Forbush decreases  observed at ground levels.
 \label{fig:vp-labels}}
\end{figure}

An interesting problem is to establish the origin of the hysteresis effect
reported by AMS02 comparing, in the same rigidity interval,
fluxes of protons and helium nuclei with a long  (378 days) averaging time,
and observing that, for the same helium flux, the He/$p$ ratio is larger
after solar maximum.
The same effect can be revealed comparing 
fluxes (or modulation potentials) of either protons or helium nuclei,
at rigidities of order 1~GV and 5~GV, and observing that 
the spectral shapes are different  before and after the solar maximum of 2014,
and for equal  flux at the lower rigidity, the flux at the higher rigidity
is larger (by approximately 4\%) after solar maximum (see Fig.~\ref{fig:corr_particles}).

Establishing the origin of this effect is not easy.
One can notice that protons and helium nuclei arrive at the
Earth mainly from the heliospheric equator before solar maximum,
and mainly from the heliospheric poles after maximum, suggesting
that the difference in
modulation could follow from this fact.
However, in conflict with the hypothesis,  one observes 
a very similar effect  (a larger flux at the higher rigidity)
for electrons, that have the opposite behaviour,  arriving at the Earth from
the poles before maximum and from the equator after maximum, 
so that the  propagation effects should be reversed.
An alternative explanations is that 
the before/after maximum asymmetry is generated by a difference in 
a ``lag effect'' of the modulations when the (time averaged) solar activity
is increasing or decreasing, and in this case one should observe the same
effect in different solar cycles.
Another possibility is that the asymmetry is the cumulative
effect of the distortions generated by solar activity events
in the early and late parts of solar cycle 24. In this case
the average effect  could be different during different solar cycles.

\vspace{0.35 cm}
In this paper we have not addressed the problem
of constructing a model of CR propagation in the heliosphere
capable of generating modulations of different shape at different times,
based in information about the state (and history) of the heliosphere.
We have however developed a preliminary step, constructing a
``minimal'' parametrization for the shape of the CR spectra at the Earth based
on a generalization of the FFA model with a rigidity dependent potential,
determined by its values at two arbitrary rigidities
(chosen as 1~GV and 10~GV here). This model allows to describe in very compact
way the differences in shape between spectra measured at different times.

Using this model, we have verified that the modulations
of protons, helium nuclei and positrons are in good approximation equal, with 
mass dependent effects smaller than few percent also at rigidities below  1~GV.
Our phenomenological model for the description of solar modulations
also suggests the intriguing result that in a broad rigidity range ([0.1,100]~GV for
$p$ and He, and [0.3,10]~GV for $e^\mp$) the LIS spectra can be well described by
a very simple form: an exact power law in rigidity
modified by an approximately constant energy loss (of order 0.3~GeV for protons and
helium nuclei, 1.1~GeV for electrons, and 0.18 for positrons).

The construction of a model that can successfully predict the
time dependence of the CR spectra at the Earth on the basis of information
about the heliosphere remains a challenging task, 
necessary to validate the reconstruction of the cosmic ray interstellar spectra.

\clearpage

\clearpage

\clearpage

\end{document}